\documentclass[a4paper,fleqn,usenatbib]{mnras}

\usepackage[T1]{fontenc}
\usepackage{ae,aecompl}

\usepackage{graphicx}	
\usepackage{amsmath}	
\usepackage{amssymb}	
\usepackage[caption=false]{subfig}
\usepackage{breakurl}

\title[A370 and MUSE]{Lens Modeling Abell 370: Crowning the Final Frontier Field with MUSE}

\author[D. J. Lagattuta et al.]
{David J. Lagattuta,$^{1}$\thanks{E-mail:david-james.lagattuta@univ-lyon1.fr}
  Johan Richard,$^{1}$
  Benjamin Cl\'{e}ment,$^{1}$
  Guillaume Mahler,$^{1}$
  \newauthor
  Vera Patr\'{i}cio,$^{1}$
  Roser Pell\'{o},$^{2,3}$
  Genevi\`{e}ve Soucail,$^{2,3}$
  Kasper B. Schmidt,$^{4}$
  \newauthor
  Lutz Wisotzki,$^{4}$
  Johany Martinez,$^{1}$
  and David Bina$^{2,3}$\\
$^{1}$Univ Lyon, Univ Lyon1, Ens de Lyon, CNRS, Centre de Recherche Astrophysique de Lyon UMR5574, F-69230, Saint-Genis-Laval, France\\
$^{2}$IRAP, Institut de Recherche en Astrophysique et Plan\'{e}tologie, CNRS, 14, avenue Edouard Belin, F-31400 Toulouse, France\\
$^{3}$Universit\'{e} de Toulouse, UPS-OMP, Toulouse, France\\
$^{4}$Leibniz-Institut f\"{u}r Astrophysik Potsdam (AIP), An der Sternwarte 16, 14482, Potsdam, Germany
}

\date{Accepted XXX. Received YYY; in original form ZZZ}

\pubyear{2016}

\begin{document}
\label{firstpage}
\pagerange{\pageref{firstpage}--\pageref{lastpage}}
\maketitle

\begin{abstract}

We present a strong lensing analysis on the massive cluster Abell 370
(A370; $z = 0.375$), using a combination of deep multi-band Hubble
Space Telescope (HST) imaging and Multi-Unit Spectroscopic Explorer
(MUSE) spectroscopy.  From only two hours of MUSE data, we are able to
measure 120 redshifts in the Southern BCG area, including several
multiply-imaged lens systems.  In total, we increase the number of
multiply-imaged systems with a secure redshift from 4 to 15, nine of
which are newly discovered.  Of these, eight are located at $z > 3$,
greatly extending the redshift range of spectroscopically-confirmed
systems over previous work.  Using these systems as constraints, we
update a parametric lens model of A370, probing the mass distribution
from cluster to galaxy scales.  Overall, we find that a model with
only two cluster-scale dark matter halos (one for each BCG) does a
poor job of fitting these new image constraints.  Instead, two
additional mass clumps -- a central ``bar'' of mass located between
the BCGs, and another clump located within a ``crown'' of galaxies in
the Northern part of the cluster field -- provide significant
improvements to the fit.  Additional physical evidence suggests these
clumps are indeed real features of the system, but with relatively few
image constraints in the crown region, this claim is difficult to
evaluate from a modeling perspective.  Additional MUSE observations of
A370 covering the entire strong-lensing region will greatly help these
efforts, further improving our understanding of this intriguing
cluster.

\end{abstract}

\begin{keywords}
gravitational lensing: strong -- galaxies: clusters: individual: Abell 370 -- techniques: imaging spectroscopy -- dark matter -- galaxies: high redshift
\end{keywords}



\section{Introduction}
\label{Intro}

Galaxy clusters acting as strong gravitational lenses are powerful
astrophysical laboratories.  The lensed background sources constrain
the cluster mass model, providing insight into the mass environments
of the densest regions of the Universe
\citep{bra08a,bra08b,hsu13,mas15}.  At the same time, the cluster
magnifies these sources, making them larger, brighter, and better
resolved.  This allows for more detailed studies of faint, low-mass,
high-redshift galaxies, opening a window into the early Universe
\citep{ebe09,sha12,mon14,pat16}.  Over the past decade, cluster
lensing science has significantly expanded, and hundreds of new lensed
systems have been discovered \citep[e.g.,][]{zit11,jau15,hoa16,kaw16}.  This
is largely thanks to improved modeling techniques (such as
\citealt{jul07} and \citealt{ogu10}) and the efforts of deep,
high-resolution imaging campaigns \citep[e.g.,][]{pos12,sch14,lot16}, driven
primarily by the \textit{Hubble Space Telescope} (HST).

But while these imaging campaigns are able to identify and resolve
background sources like never before, imaging alone does not provide
high-precision ($\Delta z < 0.01$) redshift information, a crucial
component in interpreting models and deriving their physical values.
While some spectroscopic campaigns are underway
\citep[e.g.,][]{ros14,tre15}, acquiring spectra of lensed systems has,
to date, largely been a long, costly, and inefficient process.  With
the advent of Integral Field Unit (IFU) spectroscopy, however, this
situation is rapidly changing.  Leading the way in these efforts is
the Multi-Unit Spectroscopic Explorer (MUSE; \citealt{bac10}) a
wide-field IFU on the Very Large Telescope (VLT) in Chile.  With a
large $1 \time 1$ arcmin$^2$ field of view and high sensitivity
between 4800 and 9300 \AA, MUSE is an incredibly efficient redshift
engine, providing several hundred redshifts between $z = 0$ and $z =
6$ in only a few hours of exposure time \citep[e.g.][]{bac15}.  As an
IFU, MUSE is also well-suited for blind redshift surveys, able to
detect emission lines from continuum-free sources without
pre-selecting a redshift range.  These objects (which are often missed
in traditional broad-band imaging campaigns) include high-redshift,
lensed Lyman-$\alpha$ emitters.  This has been, unsurprisingly, a boon
to the lensing community, and some studies have already begun to take
advantage of MUSE data \citep[e.g.,][]{kar15,ric15,bin16,jau16,cam16}.

In this work, we present new MUSE data of the strong lensing cluster
Abell 370 (A370; \citealt{abe58}). A370 is a massive cluster ($M_{\rm
  vir} = 2.3\times10^{15} M_\odot$; \citealt{ume11}) at redshift $z =
0.375$ \citep{mel88} with historical significance to lensing: it
contains one of the first known ``Giant Luminous Arc'' features
\citep{lyn86}, which later became the first
spectroscopically-confirmed giant-arc lens system
\citep{sou87,sou88,lyn89}.  Thanks to this discovery, and the
identification of other lensed features \citep{for88,mel91,kne93},
A370 became an important benchmark in the early days of lens modeling
\citep[e.g.,][]{kne93,sma96,bez99a,bez99b}. However, after an initial
period of activity, interest in A370 slowly waned; there was little
spectroscopic follow-up on the cluster, and it was not selected to be
a part of massive cluster surveys such as MACS \citep{ebe01} or CLASH
\citep{pos12}, in spite of its high X-ray luminosity ($L_x =
1.1\times10^{45}$ erg s$^{-1}$; \citealt{mor07}).

In 2010, a newly-refurbished HST observed A370 to test its
deep-imaging capabilities, generating renewed interest.  This led to a
new lens model \citep{ric10}, additional weak-lensing analyses
\citep{med11,ume11}, and its eventual selection as one of the Hubble
Frontier Fields (HFF) clusters \citep{lot16}.  Several groups have
since modeled the cluster (see the HFF
archive\footnote{\url{https://archive.stsci.edu/pub/hlsp/frontier/abell370/models/}}
and also \citealt{ric14} and \citealt{jon14}), but (until recently)
only three lensed systems have had a secure redshift, all with $z <
1.2$ (see section \ref{OldSys} for details.)  As a result, these
models are only able to probe the critical line region of the cluster
out to small radii.  While recent work by \citet{die16} has increased
the number of secure-redshift systems to five, the highest of these is
still $z < 2$.  Here, we extend the redshift range of
spectroscopically-confirmed systems even further and present
newly-discovered systems as well.  By combining the MUSE spectroscopy
with the current HFF data, we can therefore improve on existing lens
models, probing the mass distribution out to larger scales and with
much higher precision.  In this way, we can improve our view of this
massive cluster in conjunction with the final HFF data release.

Overall, this work is organized as follows: In Section 2, we describe
the MUSE and HFF data and the data reduction processes we used.  In
Section 3, we describe how we extract MUSE redshifts and present the
A370 redshift catalog, paying particular attention to redshifts of the
multiply-imaged galaxies.  Using these redshifts, we construct a new
mass model, which we present in Section 4.  We discuss the results of
the mass modeling and make predictions for future work in Section 5.
Finally, we briefly conclude in Section 6.  Throughout this paper, we
assume a standard cosmological model with $\Omega_M = 0.3$,
$\Omega_{\Lambda} = 0.7$, and $H_0$ = 70 km s$^{-1}$ Mpc$^{-1}$.
Using this model, one arcsecond covers a physical distance of 5.162
kpc at the A370 redshift ($z = 0.375$).  All magnitudes are measured
using the AB system.

\section{Data}
\label{Data}

Lensing is a three-dimensional effect, sensitive to both the
transverse and line-of-sight distances between objects.  Therefore, we
need a combination of imaging data (to identify lensed objects and
measure their apparent positions) and redshift data (to measure radial
distances) to construct an accurate lens model of A370.

\subsection{HST Imaging}
\label{HST}

HST imaging data of A370 were taken as part of the Hubble Frontier
Fields (HFF) Program, and are publicly available on the HFF
website\footnote{\url{http://www.stsci.edu/hst/campaigns/frontier-fields/}}. While
the imaging campaign for A370 is now complete, the Epoch 1 (v1.0)
mosaics that we use in this work consist of deep Advanced Camera for
Surveys (ACS; \citealt{for03}) data in three optical bands, F435W,
F606W, and F814W (ID: 13504, PI: J.\ Lotz).  These are supplemented by
shallower F814W imaging from archival programs: (ID: 11507, PI:
K. Noll) and (ID: 11591, PI J.-P. Kneib).  Additionally, shallow,
archival, Wide-Field Camera 3 (WFC3; \citealt{kim08}) data is
available in three bands: F105W (ID: 13459, PI: T. Treu), F140W (ID:
11108, PI: E. Hu; ID: 13459, PI: T. Treu), and F160W (ID: 11591, PI:
J.-P. Kneib; ID: 14216, PI: R.\ Kirshner), and a shallow, pilot HFF
exposure, also in the F140W band.

Individual exposures are reduced and stacked by the HFF team directly,
using the standard Pyraf/STSDAS pipeline.  Additionally, the ACS
mosaics are further corrected with a ``self-calibration'' technique to
eliminate Charge Transfer Inefficiency (CTI) effects and low-level
pixel
noise\footnote{\url{http://www.stsci.edu/hst/acs/software/Selfcal}}
(J. Anderson, in prep).  The full reduction procedure is described in
the A370 HFF data
archive\footnote{\burl{https://archive.stsci.edu/pub/hlsp/frontier/abell370/images/hst/v1.0-epoch1/hlsp\_frontier\_hst\_acs-00\_abell370\_v1.0-epoch1\_readme.pdf}}.

After combining all exposures, total integration times are 51400,
25316, 130643, 2335, 12894, and 9647 seconds for the F435W, F606W,
F814W, F105W, F140W, and F160W bands, respectively.  This is
equivalent to 20, 10, 52, 1, 5, and 4 HST orbits.  Collectively, this
represents one of the deepest, highest resolution imaging of A370 ever
taken, and corresponds to average limiting magnitudes of 29.3 in the
optical bands and 27.6 in the near-IR bands.  A color image of the
cluster field consisting of the F435W, F606W, and F814W mosaics is
shown in Figure \ref{fig:HST}.

\subsection{MUSE Spectroscopy}
\label{MUSE}

\begin{figure*}
  \includegraphics[width=2\columnwidth]{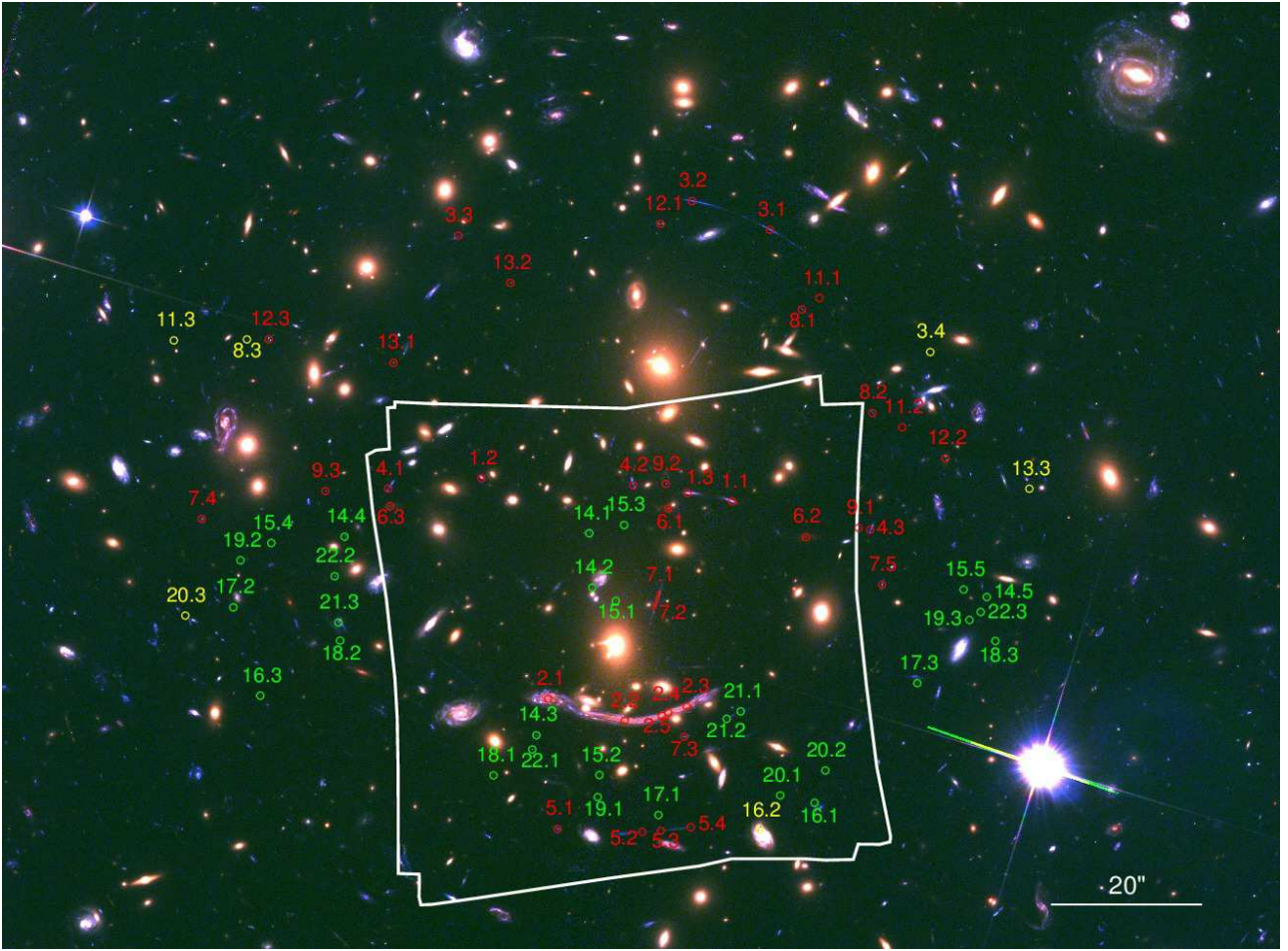}
  \caption{A color image of the A370 field of view, using the F435W,
    F606W, and F814W observations from the Hubble Frontier Field (HFF)
    project.  The region of the cluster covered by the MUSE GTO
    program is shown in white.  Additionally, the positions of the
    multiply-imaged systems used as constraints in the mass model
    (Section \ref{Model}) are shown as colored circles.
    Previously-known systems are in red, while newly-identified
    objects are in green.  Images predicted by the model but not
    detected in either imaging or spectroscopic data are shown in
    yellow.}
  \label{fig:HST}
\end{figure*}

MUSE observations of A370 were taken on UT 2014 November 20, as part
of the Guaranteed Time Observing (GTO) Program 094.A-0115(A)
(PI:Richard).  In total, we observed four 30-minute exposures in
WFM-NOAO-N mode, centered at ($\alpha =$ $2^{\rm h}$ $39^{\rm m}$
$53\fs111$, $\delta =$ $-1\degr$ $34\arcmin$ $55\farcs 77$).  We
applied a small ($\sim$ 0.5\arcsec) dither pattern between exposures
to average out systematics on the detector, and we alternated taking
exposures between PA = 0\degr\ and PA = 8\degr\ to reduce the
systematic striping pattern caused by the IFU image slicers.  The full
observational footprint can be seen in Figure \ref{fig:HST}.  Typical
seeing during the observations was 0.75\arcsec, as measured by stars
in the field.

The raw data were reduced through the MUSE pipeline
\citep{wei12,wei14} provided by ESO (version 1.2).  This pipeline
performs all basic reduction techniques: bias and flat-field
correction (using a combination of internal flats and illumination and
twilight exposures), wavelength and geometric calibration, sky
subtraction, flux calibration, and telluric correction.  The A370 data
were flux-calibrated using a combination of several standard stars
observed under photometric conditions.  After basic corrections we
align individual exposures to a common WCS with \texttt{SCAMP}
\citep{ber06}, shifting each frame relative to a reference image,
which in this case is the F814W HFF data.  We then transform the
re-aligned images into data cubes, resampling all pixels onto a common
3-dimensional grid with two spatial and one spectral axis.  The final
spectral resolution of the cubes varies from R=2000 to R=4000, with a
spectral range between 4750 and 9350 \AA.  To ensure that cubes are
properly Nyquist-sampled, we set the wavelength grid to 1.25 \AA
pixel$^{-1}$.  The final spatial resolution is 0.2\arcsec pixel$^{-1}$
in order to properly sample the PSF.

Next, we process each cube with the Zurich Atmosphere Purge
(\texttt{ZAP}; \citealt{sot16}), a software package that uses a
principal component analysis technique to remove known systematics
from the sky model, further improving sky-subtraction residuals.  To
account for changes in sky transmittance from exposure to exposure, we
compare the fluxes of stars common to each cube.  The cube with the
brightest flux values is taken as a new photometric standard, and we
scale the zeropoints of the other cubes until the fluxes agree across
all exposures.  We then merge all cubes together to create a combined
master cube.  During the merging process, we apply a 3-$\sigma$
clipping routine to reject outliers, eliminating cosmic-ray strikes
and hot pixels on the detector.  As a final step, we re-run the
\texttt{ZAP} process on the master cube in order to eliminate
low-level sky residuals that can only be seen in the improved
signal-to-noise ratio of the combined data.

\section{Redshift Measurement and Spectroscopic Catalog}
\label{Redshifts}

After reducing the data, we probe the MUSE cube for redshifts using
two complementary techniques: 1.) an automated emission line detection
program, and 2.) a customized, interactive data visualization tool
designed to extract MUSE aperture spectra by hand.

The automated program,
MUSELET\footnote{\url{http://mpdaf.readthedocs.io/en/latest/muselet.html}},
is included as part of the Muse Python Data Analysis Framework
(MPDAF\footnote{\url{https://git-cral.univ-lyon1.fr/MUSE/mpdaf}})
version 2.0.  It first creates narrow-band images of the data,
averaging the flux over a narrow emission wavelength range and then
subtracting a local continuum.  For the A370 cube, we choose an
emission window of 6.25 \AA\ (spanning five MUSE wavelength slices),
with each window centered on one of the original wavelength planes.
The continuum is created by averaging two 25 \AA\ slices, immediately
blue- and red-ward of the emission window.  After creating the
narrow-band images, MUSELET uses Source Extractor \citep{ber96} to
identify emission features in each image, then merges these features
together into a final master catalog.  Emission features at different
wavelengths are considered to belong to the same source if they fall
within 0.8\arcsec\ of each other (i.e., within the seeing FWHM) in
the case of continuum-free detections, or if they fall within half of
the effective radius ($r_e$/2) of a continuum-emission object.  For
every source that has multiple emission lines identified, MUSELET fits
a redshift using a template of known spectral lines.  To ensure the
accuracy of the MUSELET fit, we manually inspect the line features of
each identified redshift before adding it to the final catalog.

\begin{figure}
  \includegraphics[width=\columnwidth]{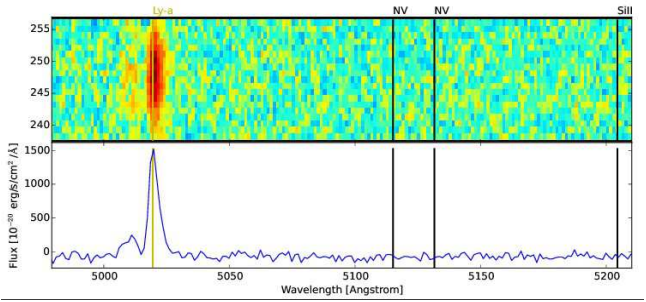}
  \caption{Sample spectrum extracted by our interactive python tool.
    Here we see Image 14.1, a newly detected Lyman-$\alpha$ emitter at
    redshift $z = 3.1309$.  The Lyman-$\alpha$ emission is clearly
    detected in the spectrum and is double-peaked.}
  \label{fig:MIDAS}
\end{figure}

In cases where an obvious galaxy appears in the HST image or MUSE cube
but is undetected (absorption galaxies) or unclassified (single
emission-line galaxies) by MUSELET, we instead use the interactive
tool.  The tool, a custom-made Python script, collapses the data cube
along the wavelength axis, creating a ``white-light'' image of the
field.  A second panel, matched to the WCS of the cube, shows a
corresponding HST image to help identify the target galaxy.  Once a
galaxy has been identified, a user can draw a (circular or
rectangular) aperture around the target object, which will extract
both a 2D and 1D spectrum from the cube.  From these spectra, the user
can then interactively fit a redshift to the galaxy, using the same
set of lines as the automatic line-finder. An example of this process
can be seen in Figure \ref{fig:MIDAS}.  As an additional check, we run
\texttt{AutoZ} \citep{bal14} on the extracted 1D spectrum.  In all
cases, we find the difference between the manual and \texttt{AutoZ}
fits to be less than $\delta z = 0.001$

\begin{figure*}
  \centerline{
    \includegraphics[width=15cm]{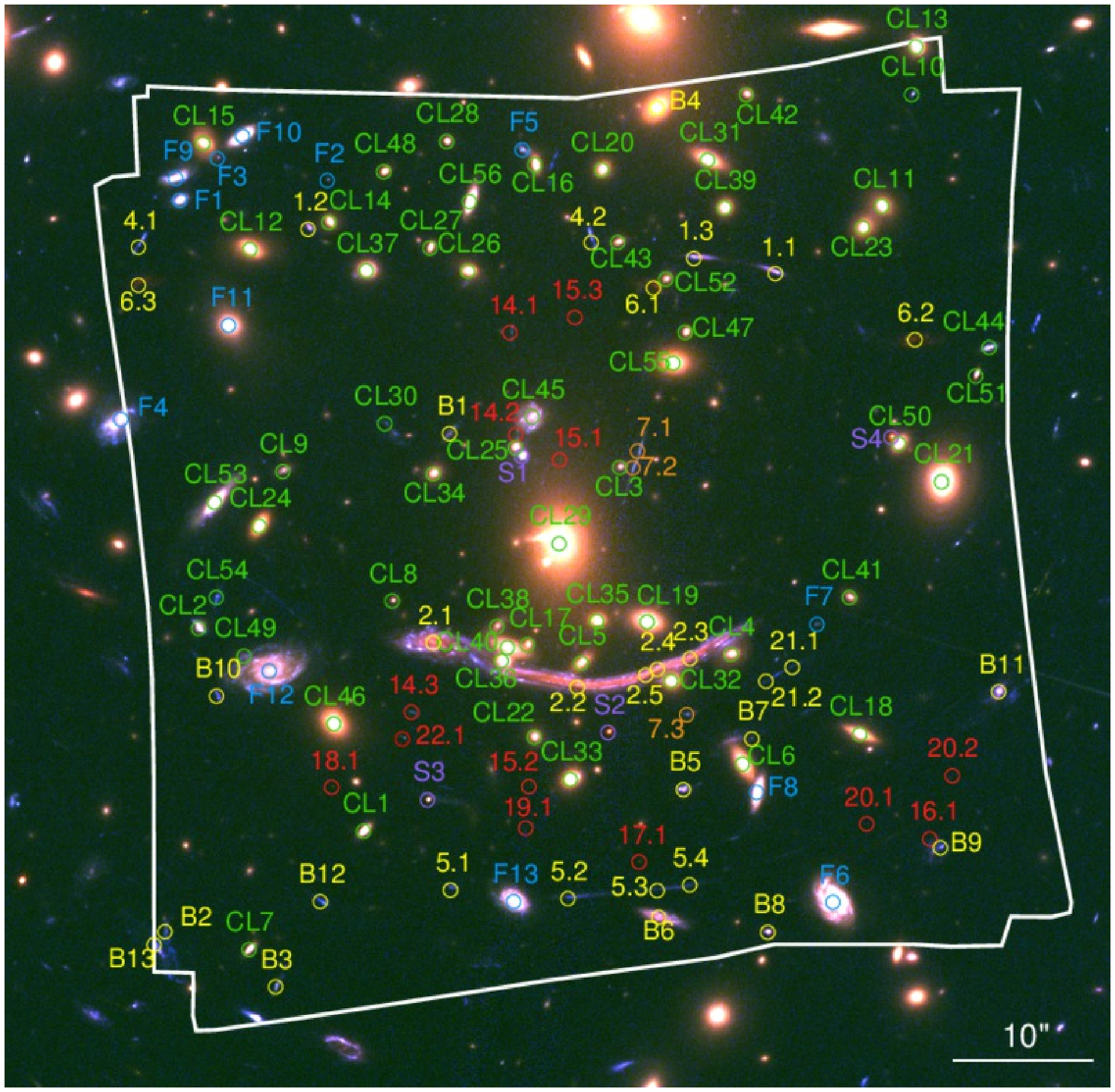}}
  \centerline{
    \includegraphics[width=\columnwidth]{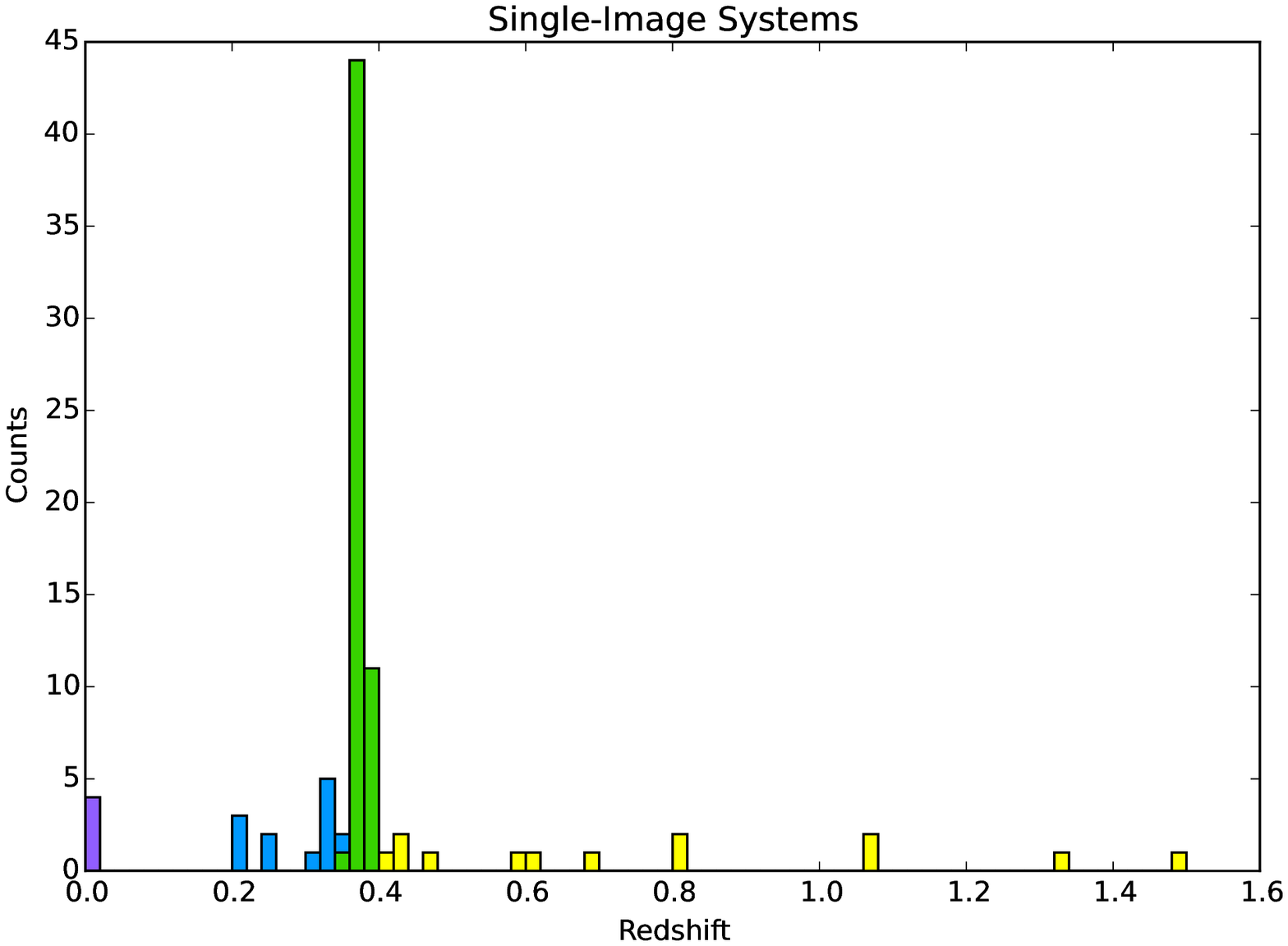}
    \includegraphics[width=\columnwidth]{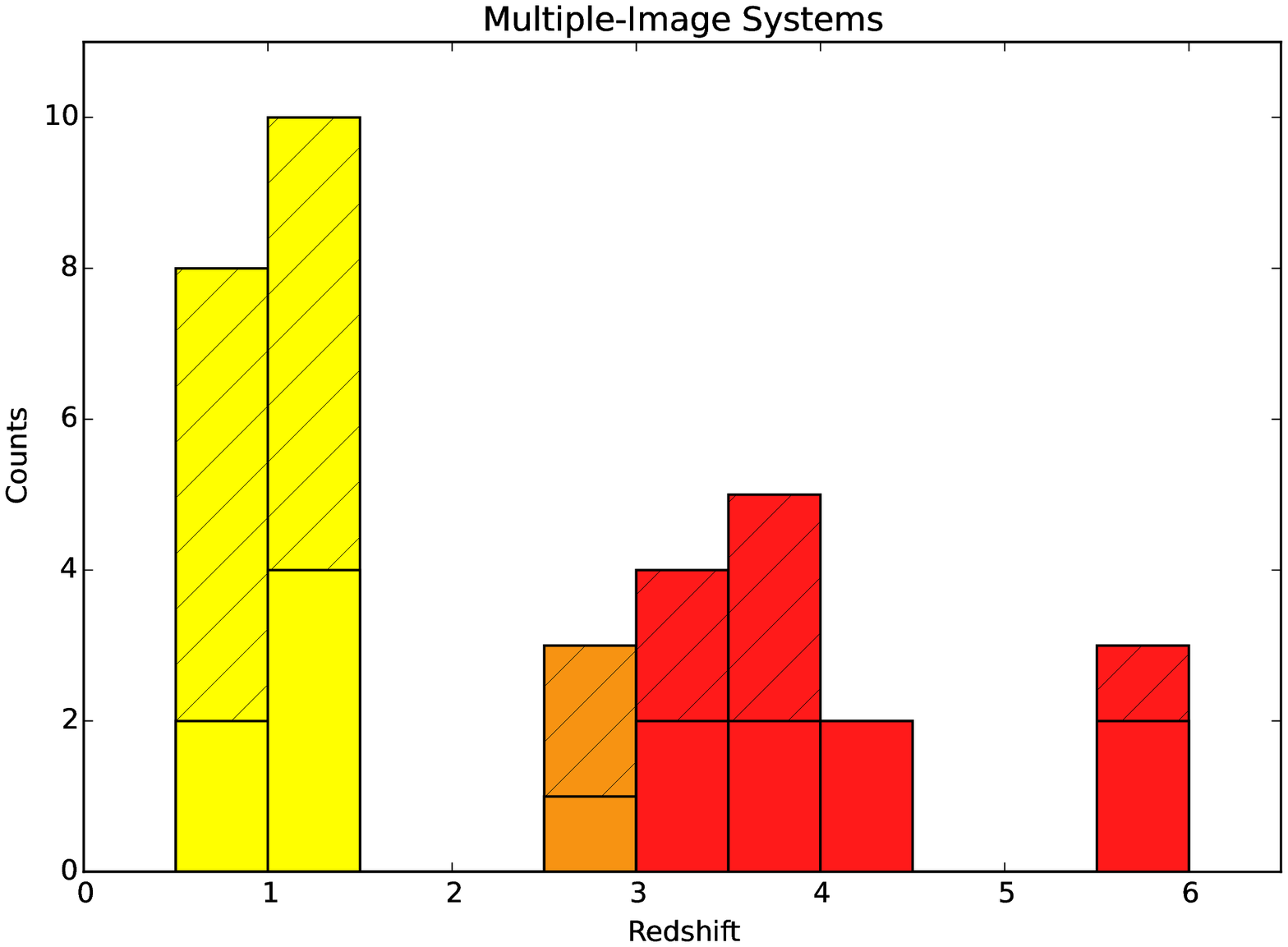}}
  \caption{{\bf: Top} Positions of all objects with a secured redshift
    in the MUSE GTO cube.  The color of each circle represents the
    object's type and redshift range, according to the following
    scheme.  Purple: star ($z = 0$), Blue: foreground galaxy ($0 < z <
    0.35$), Green: cluster member ($0.35 \leq z \leq 0.4$), Yellow:
    [\ion{O}{II}] emitter ($0.4 < z < 1.5$), Orange: \ion{C}{III}]
  emitter ($1.5 \leq z < 3.0$), Red: Ly-$\alpha$ emitter ($z \geq
  3.0$).  Object labels correspond to entries in Table
  \ref{tbl:Multi-Images} (Multiply-Imaged Systems), Table
  \ref{tbl:ClusterZ} (Cluster Members), and Table \ref{tbl:OtherZ}
  (Foreground and Background Objects.)  {\bf Bottom Left:} Redshift
  distribution of all singly-imaged systems in the cube.  The color
  scheme is the same as in the position map.  A clear overdensity of
  galaxies can be seen at the cluster redshift range, as expected.
  {\bf Bottom Right:} Similar redshift distribution, but for the
  multiply-imaged systems.  Solid color lines represent individual
  systems, while the hashed lines represent all the counterimages that
  make up these systems.}
      \label{fig:Zdist}
\end{figure*}

After running both methods on the A370 cube, we combine all of the
redshift results into a final catalog.  Overall, we securely identify
120 redshifts, consisting of multiply-imaged systems, cluster members,
foreground interlopers, background sources, and stars.  The spatial and
spectral distributions of these redshifts are shown in Figure
\ref{fig:Zdist}.

\subsection{Multiply-Imaged Systems}
\label{Multi}

Multiply-imaged background sources are particularly important in this
work, since they provide strong constraints on the lensing mass model
(see Section \ref{Model}). Using a combination of HFF imaging and MUSE
spectroscopy, we are able to identify and confirm the redshifts of
previously known multi-image systems and identify new systems as well.

\subsubsection{Previously Known Systems}
\label{OldSys}

Prior to the release of the newest HFF and MUSE data, 13 multi-image
objects were identified in the A370 field \citep{ric14} including
three with known redshifts: Systems 1 \citep{kne93}, 2 \citep{sou88},
and 6 \citep{ric14}.  A tentative guess for System 4 ($z = 1.275$)
using VLT/FORS2 grism spectroscopy was also presented in
\citep{ric14}.  Of these systems, seven have at least one image that
falls within the MUSE A370 footprint: Systems 1, 2, 4, 5, 6, 7, and 9.
Using the MUSE data, we are able to confirm the previous redshifts of
Systems 1 ($z = 0.8041$), 2 ($z = 0.7251$), and 6 ($z = 1.0633$),
refine and secure the redshift for System 4 ($z=1.2728$), and provide
new redshifts for Systems 5 ($z = 1.2775$) and 7 ($z = 2.7512$).  We
note that these new redshifts are within 2-$\sigma$ of the values
predicted in the \citet{ric14} lens model (System 4: $z = 1.34 \pm
0.03$; System 5: $z = 1.30 \pm 0.05$; System 7: $z = 4.97 \pm 1.17$),
a reasonably good agreement.  We are unable to identify any strong
features in System 9 because it is too faint for a secure measurement.
However, \citet{die16} identify [OII] and [OIII] features for the system
in HST grism spectroscopy, placing it at $z = 1.52.$ We note that this
redshift is consistent with a non-detection in MUSE, since it falls in
the ``redshift desert'' where no strong emission features appear in
the MUSE wavelength range.  Additionally, \citet{die16} also identify
faint [OIII] emission for System 3 at $z = 1.95$.  We adopt these
redshift values in this work, also.

\subsubsection{New Systems}
\label{NewSys}

\begin{figure}
  \includegraphics[width=\columnwidth]{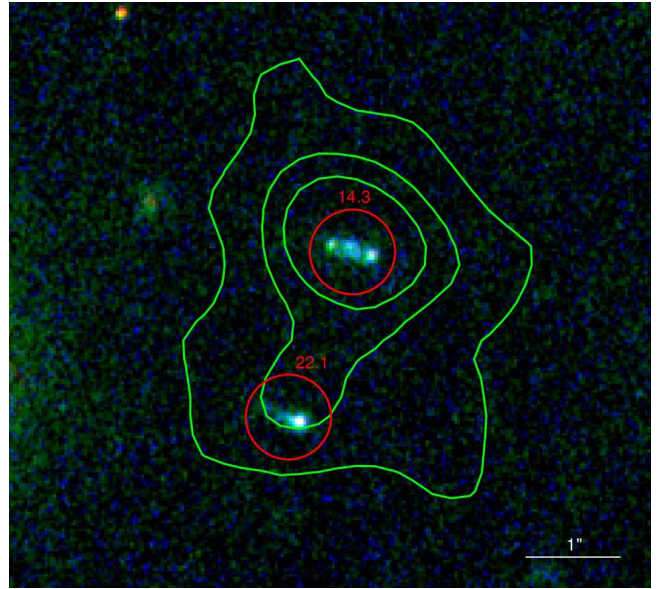}
  \caption{Close-up of the area containing Images 14.3 and 22.1, two
    newly-identified, multiply-imaged, Lyman-$\alpha$-emitting
    galaxies. While HST imaging alone shows that the two galaxies are
    in close proximity to one another and have similar colors, new
    spectroscopy from MUSE reveals that they have exactly the same
    redshift ($z = 3.1309$.)  Additionally, narrow-band imaging of the
    emission line (green contours) shows that there is a significant
    overlap between the galaxies' Lyman-$\alpha$ halos, suggesting
    that the two galaxies could be interacting.}
  \label{fig:14+22}
\end{figure}

In addition to the known objects, we also identify nine new
multiply-imaged systems in the MUSE field (Systems 14 -- 22).
\citet{die16} independently identify four of these systems: 14, 20,
21, and 22 (labeled Systems 10, 29, 26, and 18, respectively, in their
work) though without spectroscopic redshifts.  With the exception of
System 21 (identified by [OII]), these new systems are all strong
Lyman-$\alpha$ emitters, located at considerably higher redshifts ($3
< z < 6$) than those previously identified.  Systems 14 and 22 are
particularly interesting, as they are reasonably close together and
have an identical redshift ($z = 3.1309$).  This suggests that they
could be an interacting pair of galaxies, and indeed, narrow-band
imaging shows that the emission-line regions of the two are clearly
overlapping (Figure \ref{fig:14+22}.)  Careful observation of the HFF
data reveals at least some flux associated with each of the new
systems in broadband imaging, but in some cases -- especially Systems
15 and 16 -- this flux is extremely faint and contaminated by brighter
objects nearby.  Spectra of all systems in the MUSE field of view are
shown in Figure \ref{fig:redshifts}.  The full list of multiply-imaged
systems, including all identified and/or model-predicted counterimages
of a given system, is presented in Table \ref{tbl:Multi-Images}.

\begin{table*}
  \centering
  \caption{Multiply-Imaged Systems}
  \label{tbl:Multi-Images}
  \begin{tabular}[t]{lllcc}
    \hline
    ID$^{\rm ~a, b}$ & RA & Dec & $z^{\rm ~c}$ & $z$ range\\
    \hline
    1.1 & 39.967047 & -1.5769172 & 0.8041 & \\
    1.2 & 39.976273 & -1.5760558 & 0.8041 & \\
    1.3 & 39.968691 & -1.5766113 & 0.8041 & \\[2pt]
    \hline
    2.1 & 39.973825 & -1.5842290 & 0.7251 & \\
    2.2 & 39.971003 & -1.5850422 & 0.7251 & \\
    2.3 & 39.968722 & -1.5845058 & 0.7251 & \\
    2.4 & 39.969394 & -1.5847328 & 0.7251 & \\
    2.5 & 39.969630 & -1.5848508 & 0.7251 & \\[2pt]
    \hline
    3.1 & 39.965658 & -1.5668560 & 1.95 $^{\rm d}$ & \\
    3.2 & 39.968526 & -1.5657906 & 1.95 $^{\rm d}$& \\
    3.3 & 39.977293 & -1.5672022 & 1.95 $^{\rm d}$& \\
    *\it 3.4 & \it 39.959758 & \it -1.5713806 & -- & \\[2pt]
    \hline
    4.1 & 39.979704 & -1.5764364 & 1.2728 & \\
    4.2 & 39.970688 & -1.5763221 & 1.2728 & \\
    4.3 & 39.961971 & -1.5779671 & 1.2728 & \\[2pt]
    \hline
    5.1 & 39.973473 & -1.5890463 & 1.2775 & \\
    5.2 & 39.971110 & -1.5892363 & 1.2775 & \\
    5.3 & 39.969472 & -1.5890961 & 1.2775 & \\
    5.4 & 39.968580 & -1.5890045 & 1.2775 & \\[2pt]
    \hline
    6.1 & 39.969405 & -1.5771811 & 1.0633 & \\
    6.2 & 39.964334 & -1.5782307 & 1.0633 & \\
    6.3 & 39.979641 & -1.5770904 & 1.0633 & \\[2pt]
    \hline
    7.1 & 39.969788 & -1.5804299 & 2.7512 & \\
    7.2 & 39.969882 & -1.5807608 & 2.7512 & \\
    7.3 & 39.968815 & -1.5856313 & 2.7512 & \\
    *7.4 & 39.986567 & -1.5775688 & -- & \\
    *7.5 & 39.961533 & -1.5800028 & -- & \\[2pt]
    \hline
    *8.1 & 39.964485 & -1.5698065 & \{2.042\} & [0.5 -- 5.0]\\
    *8.2 & 39.961889 & -1.5736473 & \{2.042\} & \\
    \it *8.3 & \it 39.984904 & \it -1.5709139 & \it \{2.042\} & \\[2pt]
    \hline
    9.1 & 39.962402 & -1.5778911 & 1.52 $^{\rm d}$ & \\
    9.2 & 39.969486 & -1.5762654 & 1.52 $^{\rm d}$ & \\
    9.3 & 39.982022 & -1.5765337 & 1.52 $^{\rm d}$ & \\[2pt]
    \hline
    *10.1 & 39.968585 & -1.5717898 & -- & \\
    *10.2 & 39.968017 & -1.5708820 & -- & \\[2pt]
    \hline
    *11.1 & 39.963839 & -1.5693802 & \{4.667\} & [2.5 -- 10.0]\\
    *11.2 & 39.960789 & -1.5741702 & \{4.667\} & \\
    \it *11.3 & \it 39.987592 & \it -1.5709501 & \it \{4.667\} & \\[2pt]
    \hline
  \end{tabular}
  \begin{tabular}[t]{lllcc}
    \hline
    ID & RA & Dec & $z$ & $z$ range\\
    \hline    
    *12.1 & 39.969682 & -1.5666360 & \{2.858\} & [0.5 -- 5.0]\\
    *12.2 & 39.959198 & -1.5753221 & \{2.858\} & \\
    *12.3 & 39.984100 & -1.5709127 & \{2.858\} & \\[2pt]
    \hline
    *13.1 & 39.979513 & -1.5717782 & \{5.119\} & [0.5 -- 5.0]\\
    *13.2 & 39.975210 & -1.5688203 & \{5.119\} & \\
    \it*13.3 & \it 39.956112 & \it -1.5764444 & \{5.119\} & \\[2pt]
    \hline
    14.1 & 39.972309 & -1.5780910 & 3.1309 & \\
    14.2 & 39.972192 & -1.5801027 & 3.1309 & \\
    14.3 & 39.974254 & -1.5855770 & 3.1309 & \\
    *14.4 & 39.981313 & -1.5782202 & -- & \\
    *14.5 & 39.957673 & -1.5804590 & -- & \\[2pt]
    \hline
    15.1 & 39.971328 & -1.5806040 & 3.7084 & \\
    15.2 & 39.971935 & -1.5870512 & 3.7084 & \\
    15.3 & 39.971027 & -1.5777907 & 3.7084 & \\
    *15.4 & 39.984017 & -1.5784514 & -- & \\
    *15.5 & 39.958410 & -1.5793722 & -- & \\[2pt]
    \hline
    16.1 & 39.964016 & -1.5880782 & 3.7743 & \\
    \it*16.2 & \it39.966037 & \it-1.5890355 & -- & \\
    *16.3 & 39.984414 & -1.5841111 & -- & \\[2pt]
    \hline
    17.1 & 39.969758 & -1.5885333 & 4.2567 & \\
    *17.2 & 39.985403 & -1.5808406 & -- & \\
    *17.3 & 39.960235 & -1.5836508 & -- & \\[2pt]
    \hline
    18.1 & 39.975830 & -1.5870613 & 4.4296 & \\
    *18.2 & 39.981476 & -1.5820728 & -- & \\
    *18.3 & 39.957362 & -1.5820861 & -- & \\[2pt]
    \hline
    19.1 & 39.971996 & -1.5878654 & 5.6493 & \\
    *19.2 & 39.985142 & -1.5790944 & -- & \\
    *19.3 & 39.958316 & -1.5813093 & -- & \\[2pt]
    \hline
    20.1 & 39.965279 & -1.5878055 & 5.7505 & \\
    20.2 & 39.963619 & -1.5868798 & 5.7505 & \\
    *\it 20.3 & \it 39.986651 & \it -1.5812606 & -- & \\[2pt]
    \hline
    21.1 & 39.966733 & -1.5846943 & 1.2567 & \\
    21.2 & 39.967252 & -1.5849694 & 1.2567 & \\
    *21.3 & 39.981539 & -1.5814028 & -- & \\[2pt]
    \hline
    22.1 & 39.974406 & -1.5861017 & 3.1309 & \\
    *22.2 & 39.981675 & -1.5796852 & -- & \\
    *22.3 & 39.957906 & -1.5810108 & -- & \\

    \hline
  \end{tabular}
  \medskip\\
  \begin{flushleft}
    {$^{\rm a}$} Images labeled with an asterisk (*) fall outside of
    the MUSE cube and do not have secure redshifts.  We identify them
    in the HFF data, using the A370 mass model as a guide.\\[2pt]
    {$^{\rm b}$} Images in italics are predicted by the model, but no
    suitable counterimage is seen in HST data.  We do not include
    these images as model constraints, but we present them here for
    completeness.  Image 16.2, which is in the MUSE field of view but
    predicted to fall behind a bright cluster galaxy, is also included
    here, since it is undetected in MUSE.\\[2pt]
    {$^{\rm c}$} Redshifts enclosed in braces (\{\}) are fit by the
    model as free parameters.  The fit range is given in the ``$z$
    range'' column.\\[2pt]
    {$^{\rm d}$} Redshifts for these systems are taken from
    \citet{die16}
  \end{flushleft}      
  
\end{table*}

\begin{figure*}
  \centerline{
    \includegraphics[width=0.5\textwidth]{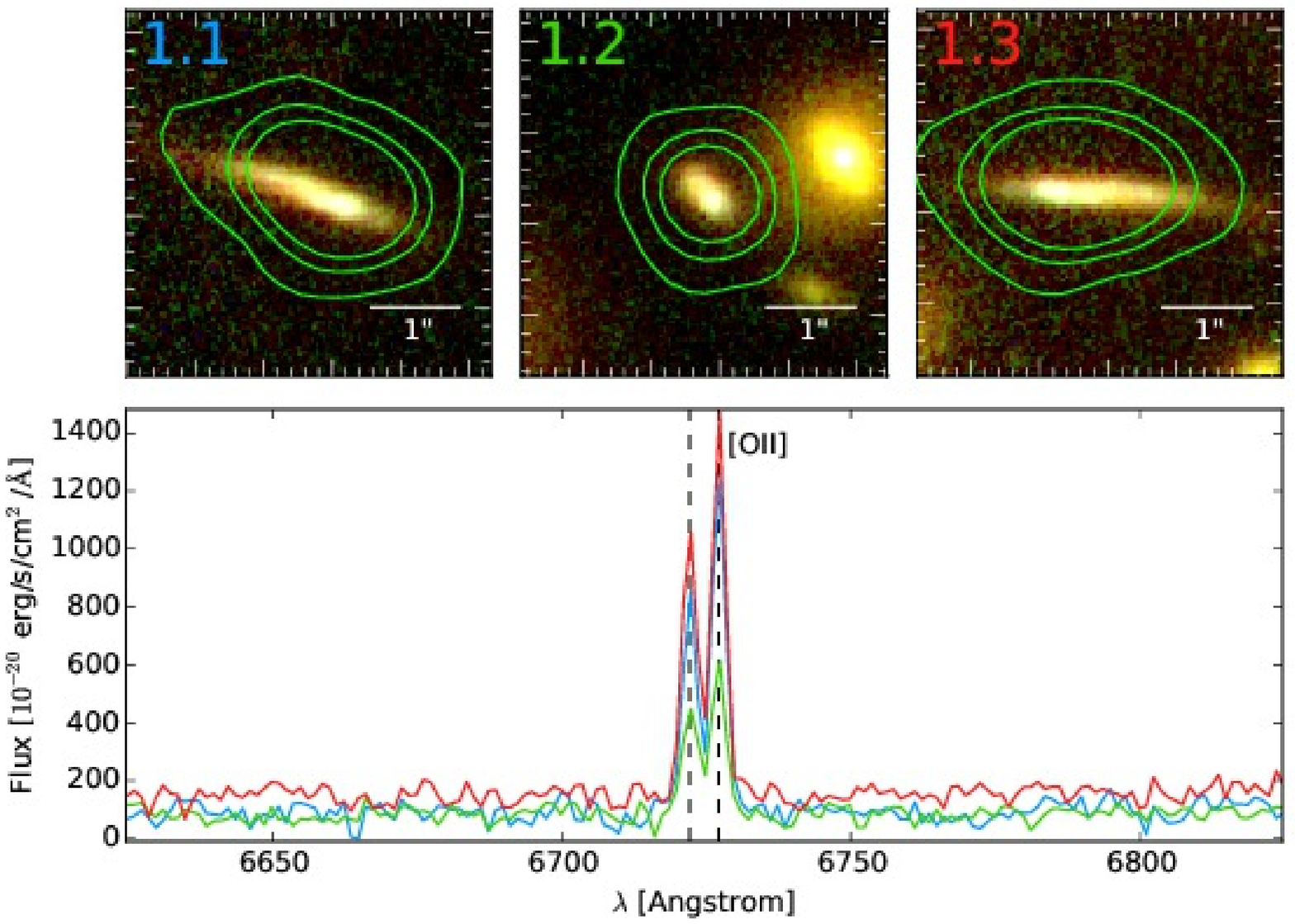}
    \includegraphics[width=0.5\textwidth]{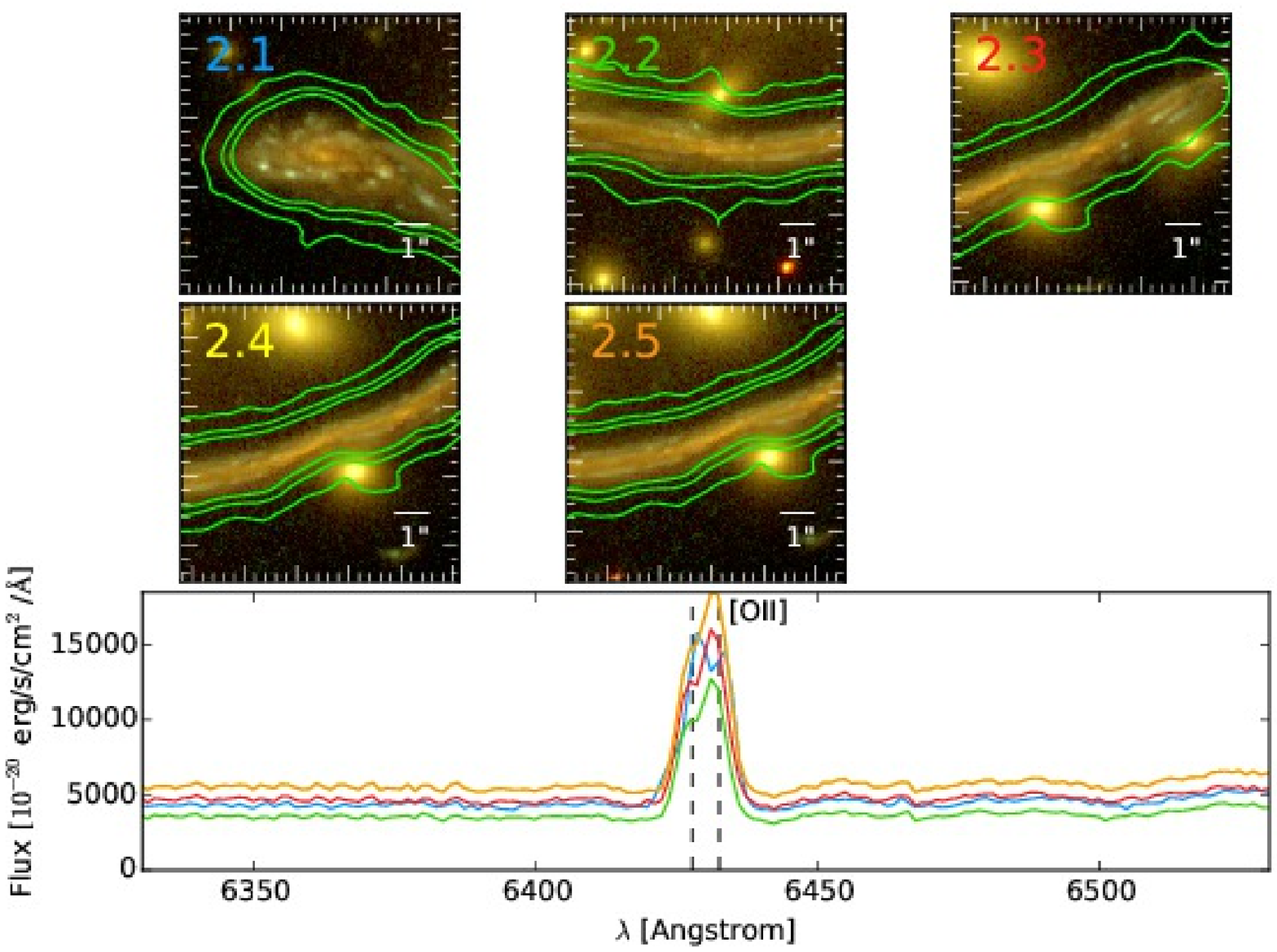}}
  \centerline{
    \includegraphics[width=0.5\textwidth]{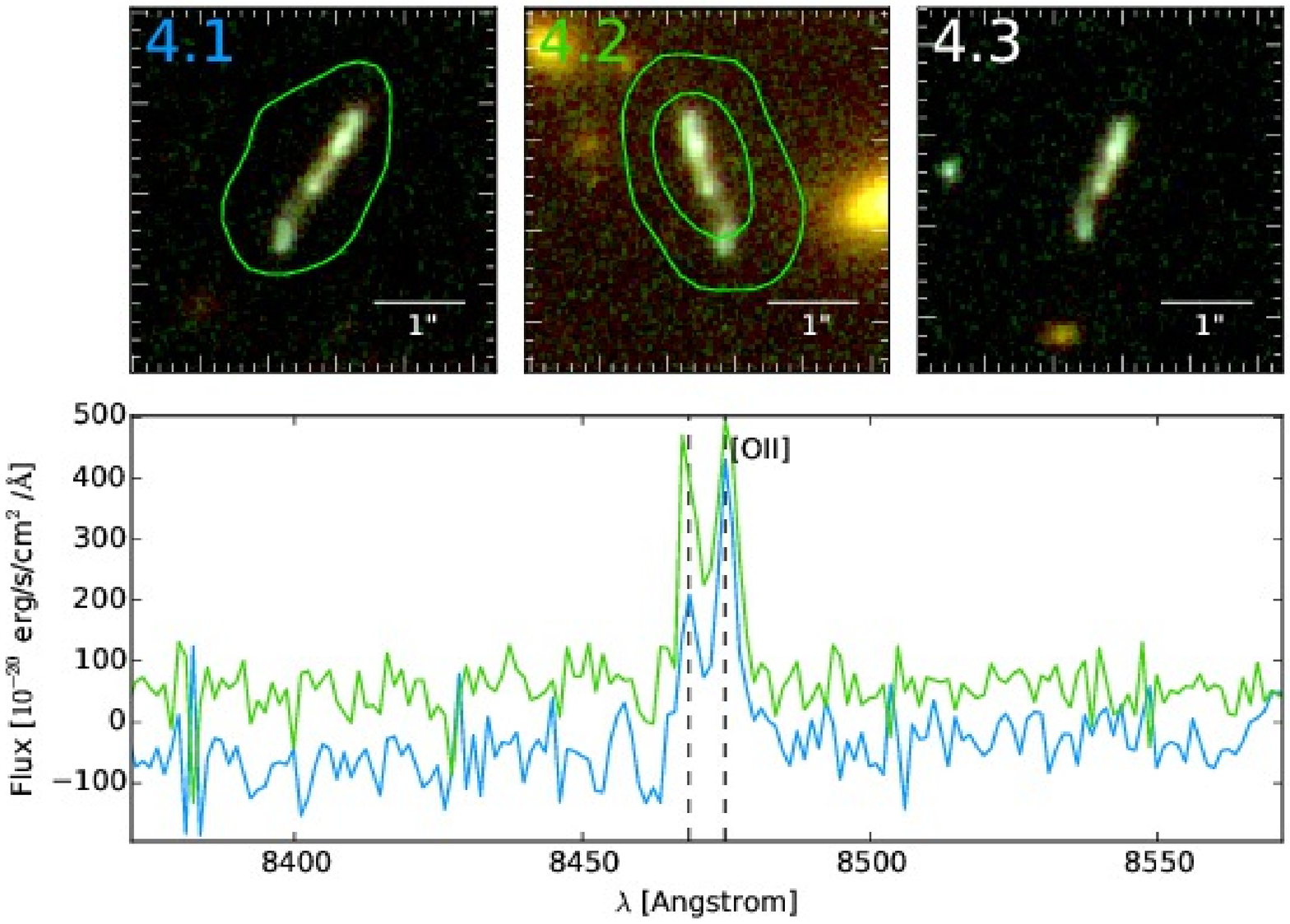}
    \includegraphics[width=0.5\textwidth]{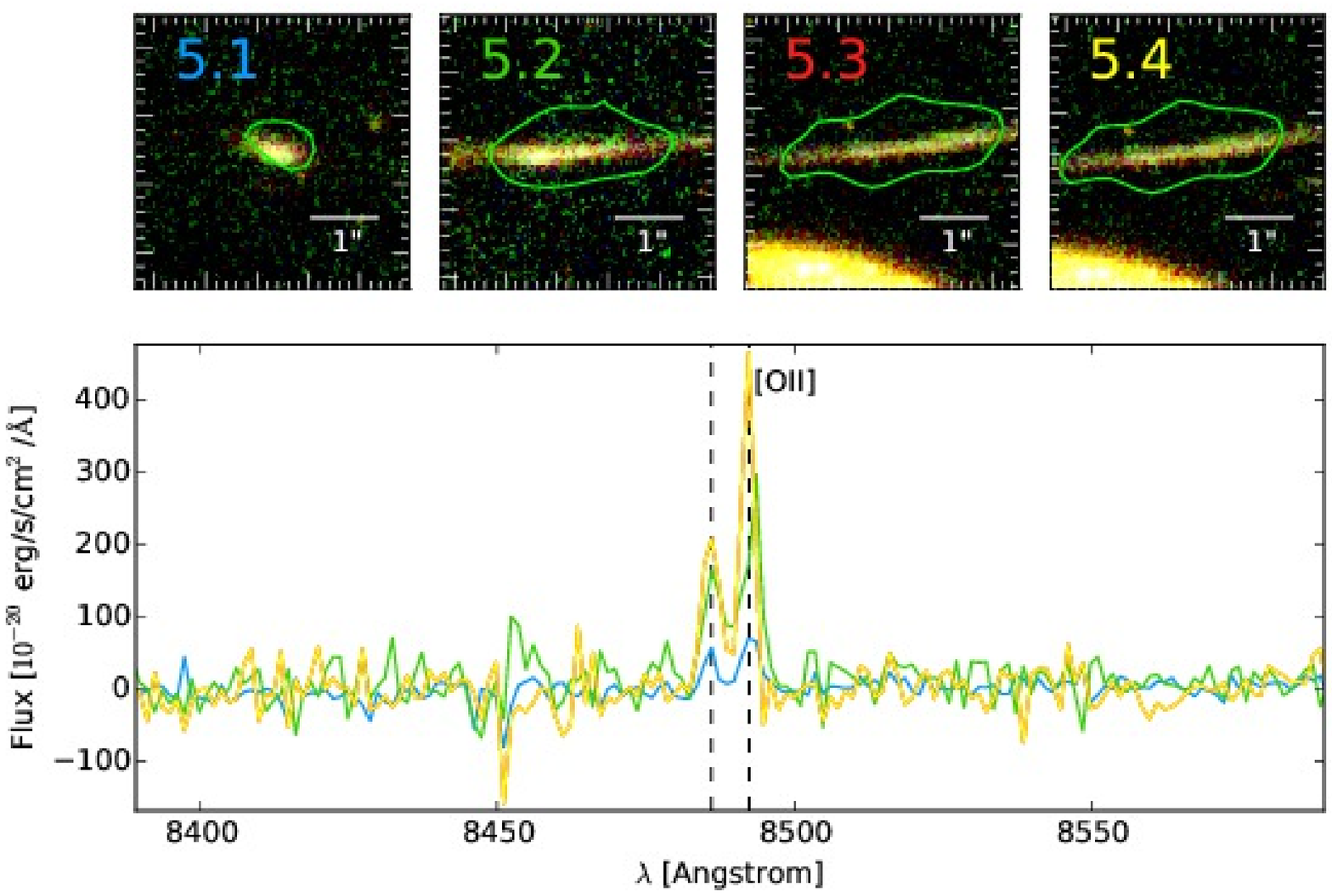}}
  \centerline{
    \includegraphics[width=0.5\textwidth]{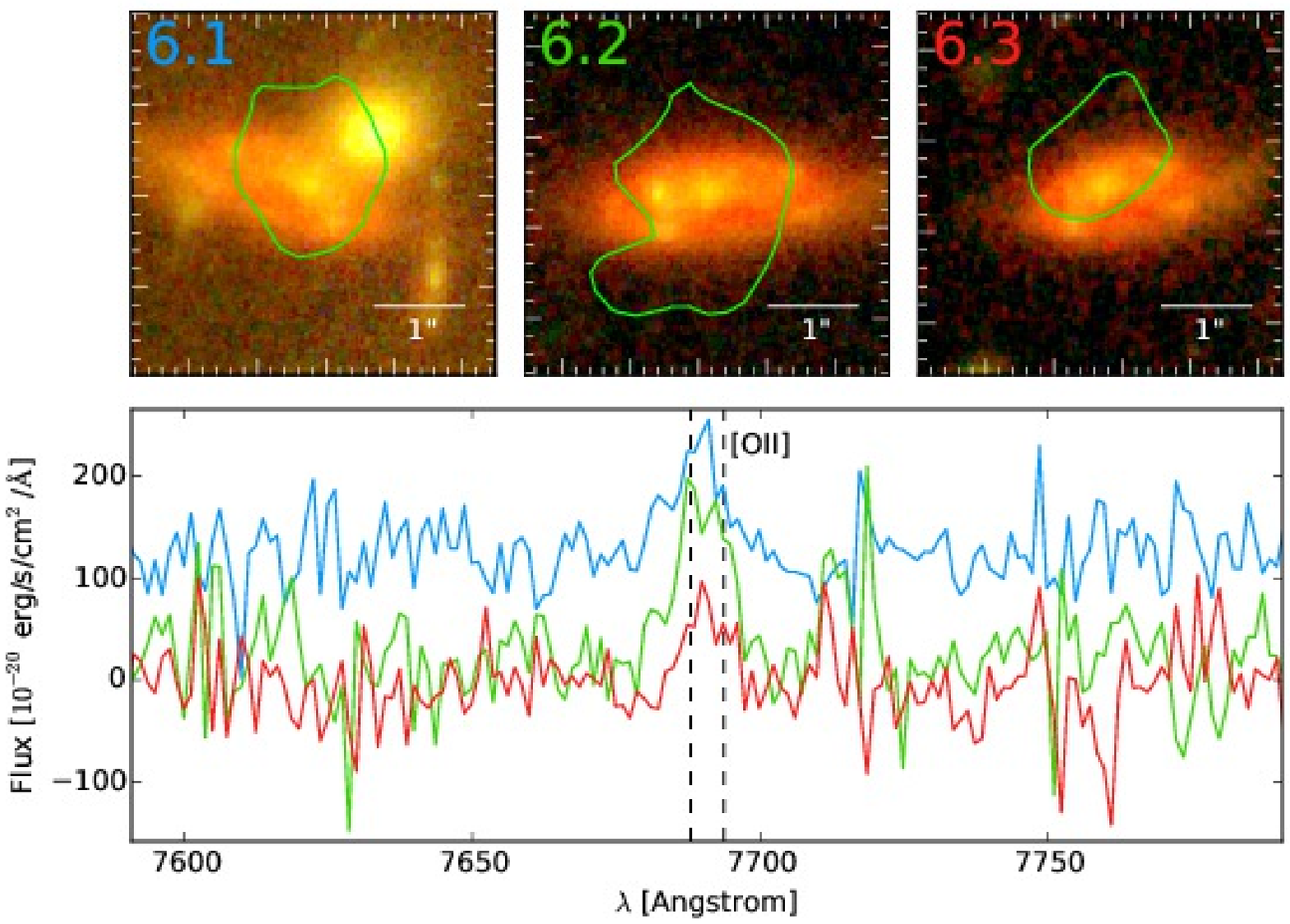}
    \includegraphics[width=0.5\textwidth]{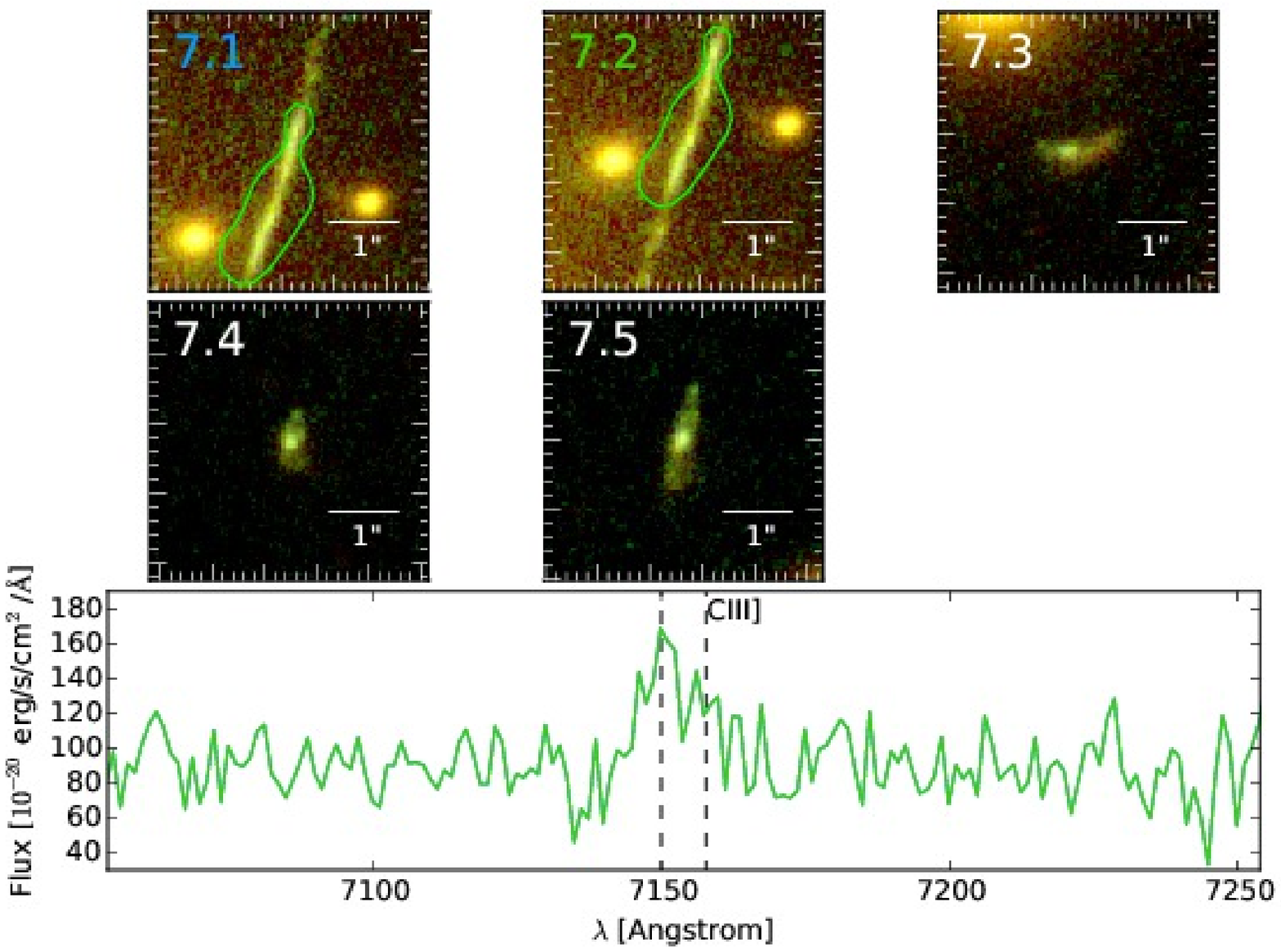}}
  \caption{Multiply-imaged systems with MUSE spectroscopy.  Individual
    members of a given system (constraints) are labeled according to
    Table \ref{tbl:Multi-Images}.  In the top panels, we show color
    images of each constraint, using the F435W/F606W/F814W HFF bands.
    The one exception is System 6, which instead uses a color scheme
    of F435W/F814W/F160W to better highlight its morphology.  Green
    contours represent the 1- 3- and 5-$\sigma$ (per pixel) levels of
    the brightest emission line seen in narrow-band imaging of the
    MUSE data cube.  Cutouts of the extracted spectra, again centered
    on the brightest emission line, are shown in the bottom panel.
    The label color of a given constraint matches the color of its
    corresponding spectrum.  Constraints that fall outside of the MUSE
    field of view, and thus have no MUSE spectroscopy, are labeled in
    white.  With the exception of System 16, which includes the
    undetected Image 16.2, only image constraints used in the A370
    mass model are displayed here.  Model predictions outside of the
    MUSE footprint where we find no obvious HST counterpart (e.g.,
    Image 20.3) are not shown.  For merging pairs of images (Images
    5.3/5.4, 7.1/7.2, and 21.1/21.2) we use a common extraction region
    for both spectra.  The combined spectrum of both images then
    appears as a single line in the plot.}
\end{figure*}

\begin{figure*}\ContinuedFloat
  \centerline{
    \includegraphics[width=0.5\textwidth]{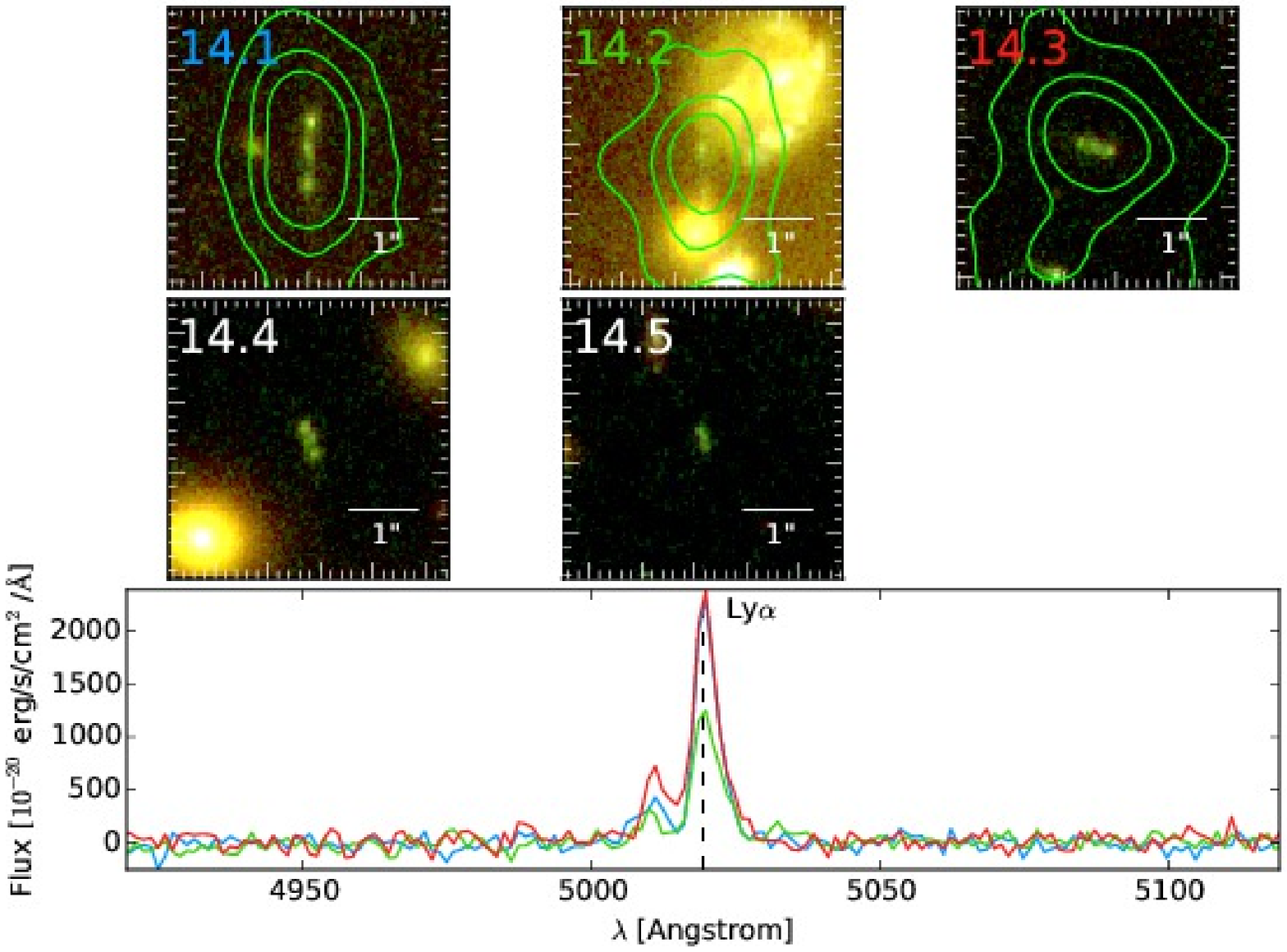}
    \includegraphics[width=0.5\textwidth]{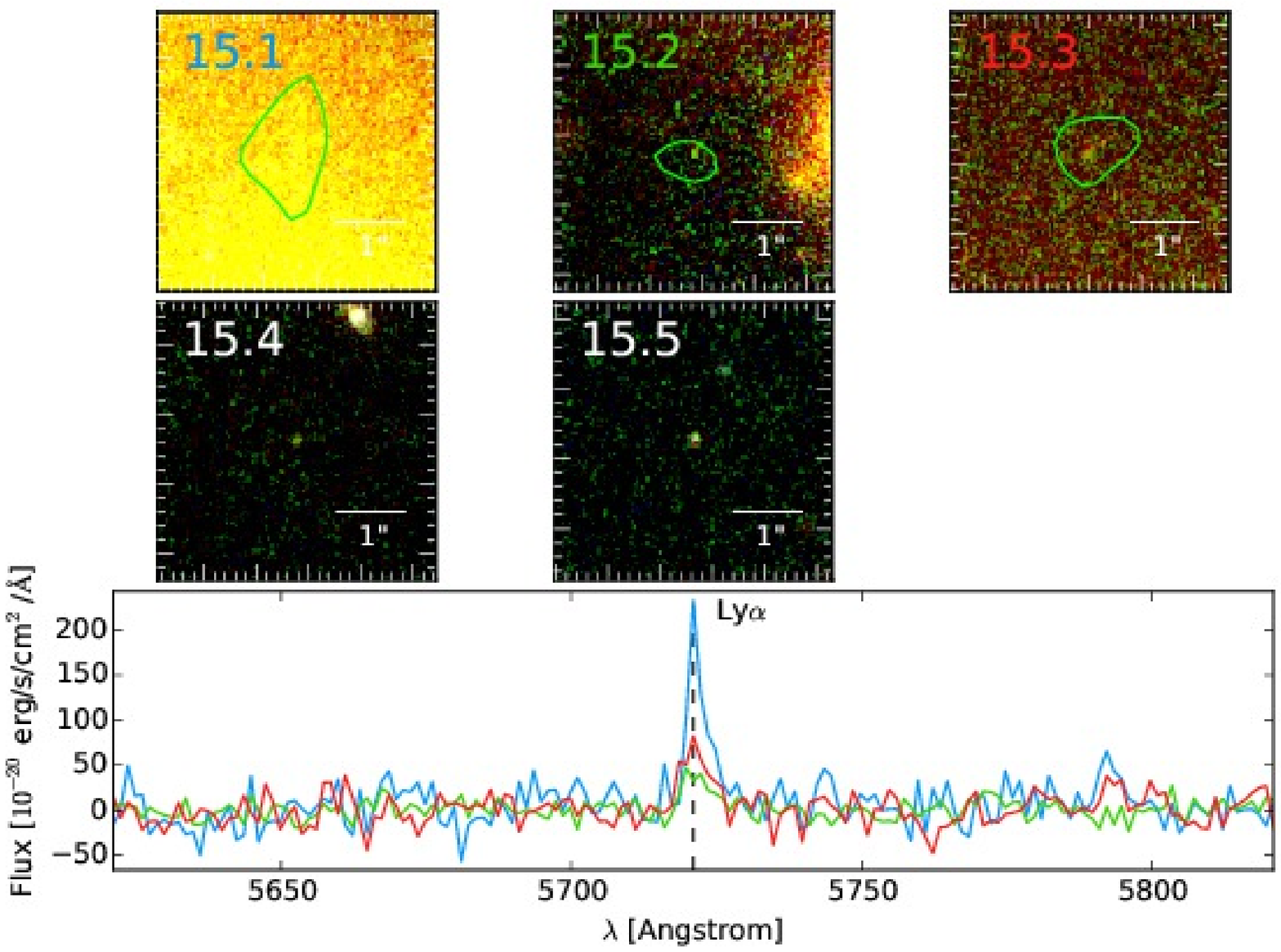}}
  \centerline{
    \includegraphics[width=0.5\textwidth]{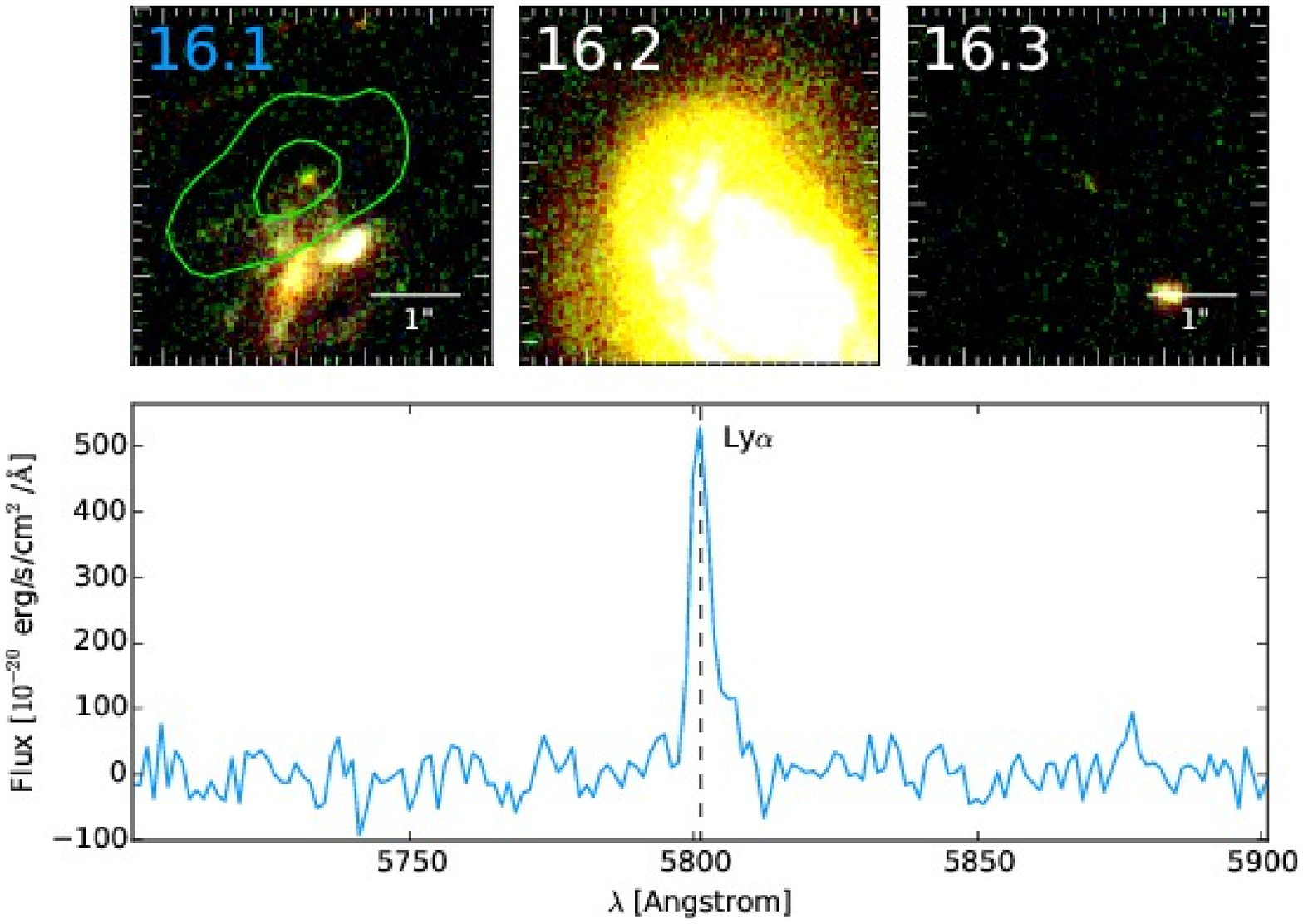}
    \includegraphics[width=0.5\textwidth]{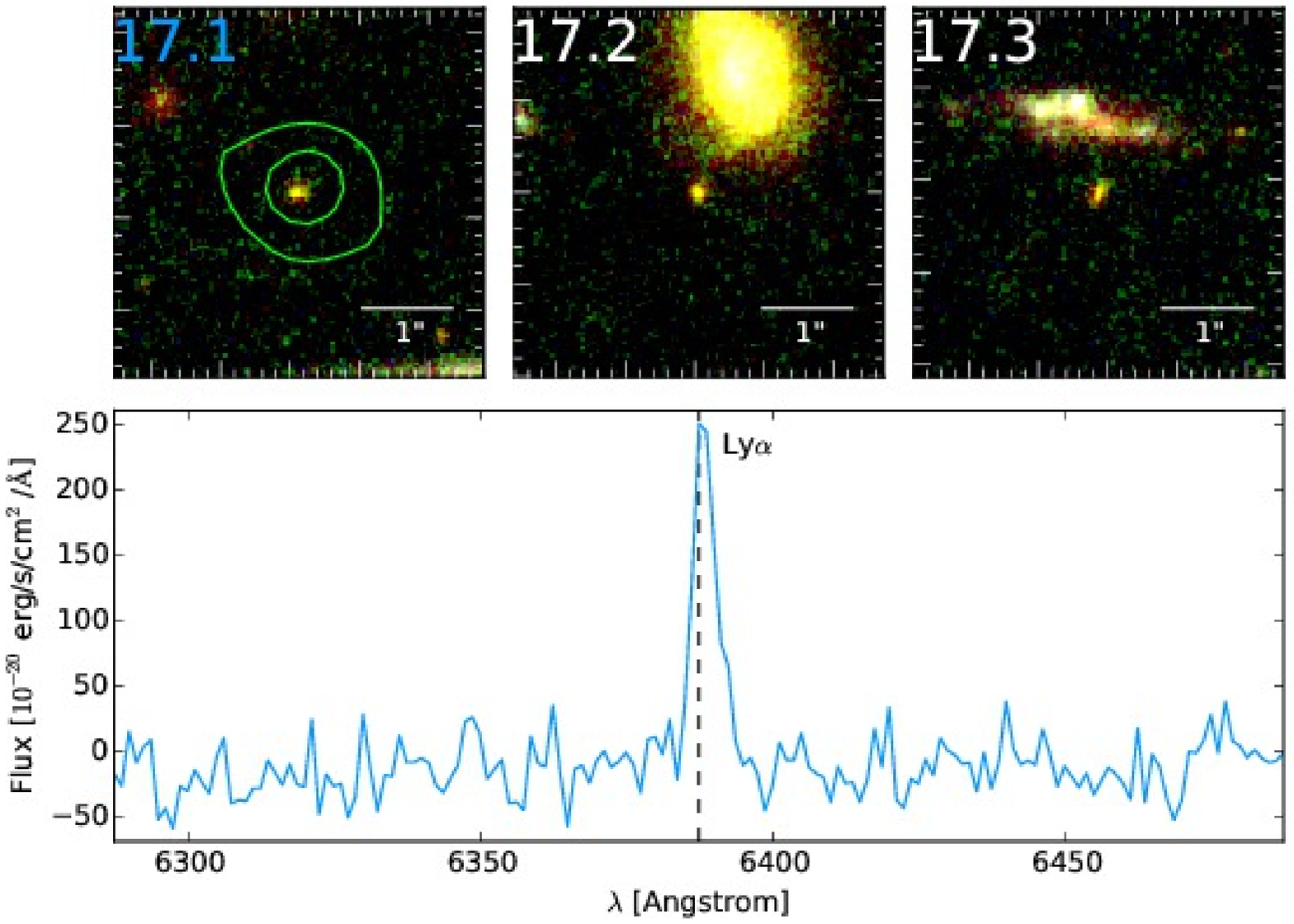}}
  \centerline{
    \includegraphics[width=0.5\textwidth]{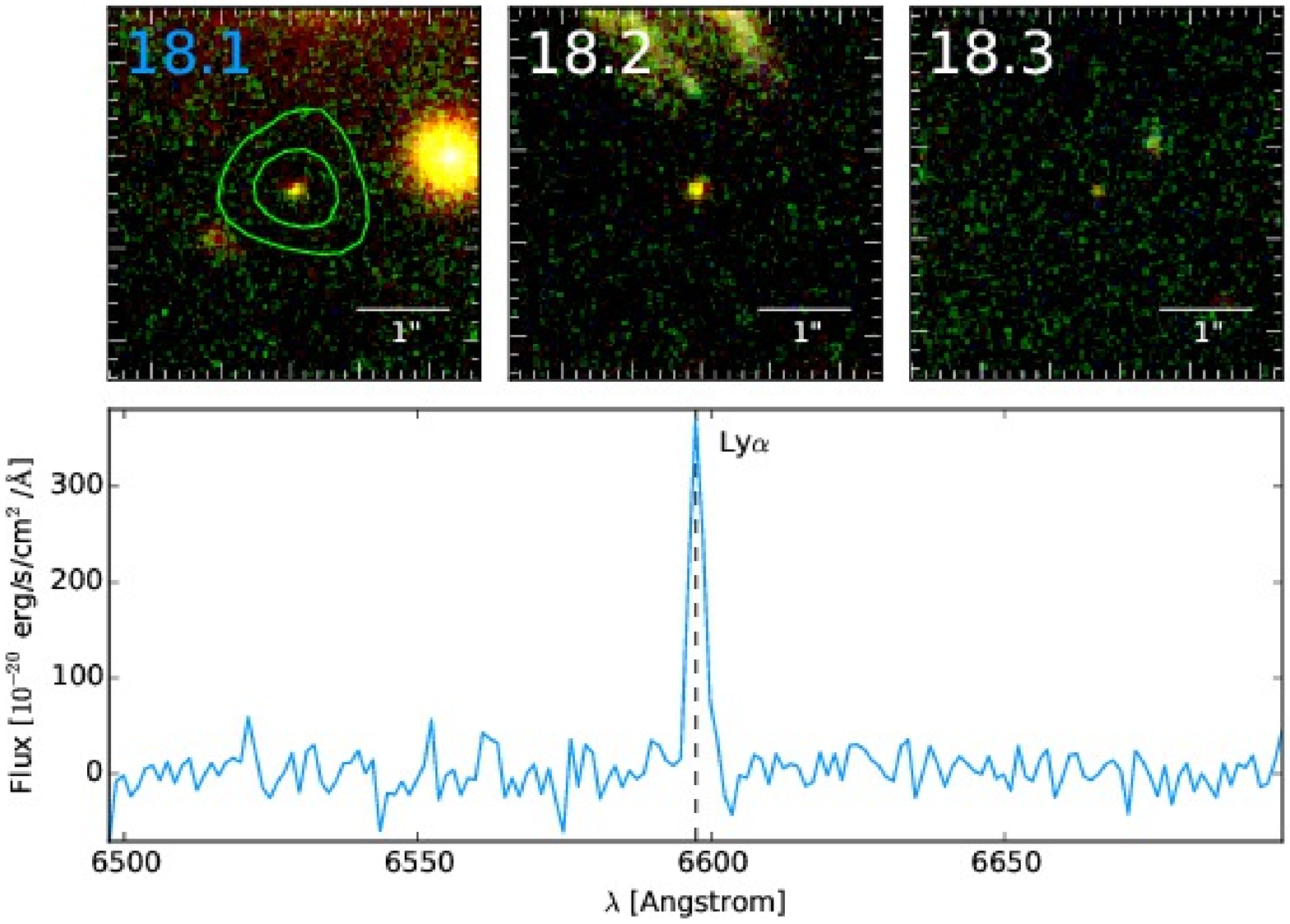}
    \includegraphics[width=0.5\textwidth]{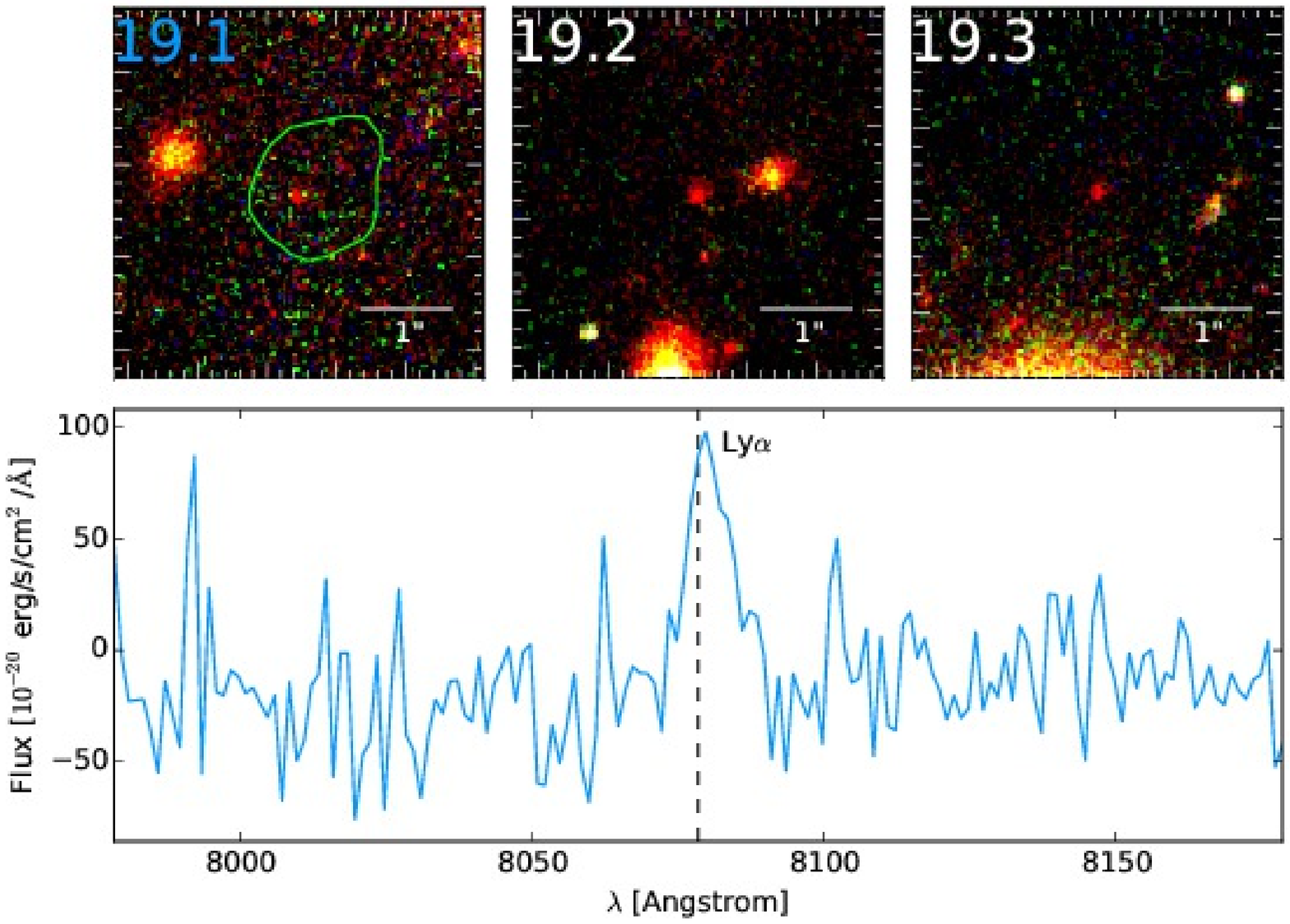}}
  \caption{(continued) New MUSE-identified systems.  The contours seen
    in Image 14.3 are identical to those in Figure \ref{fig:14+22},
    highlighting the overlap between the Lyman-$\alpha$ halos of
    Images 14.3 and 22.1.}
\end{figure*}

\begin{figure*}\ContinuedFloat
  \centerline{
    \includegraphics[width=0.5\textwidth]{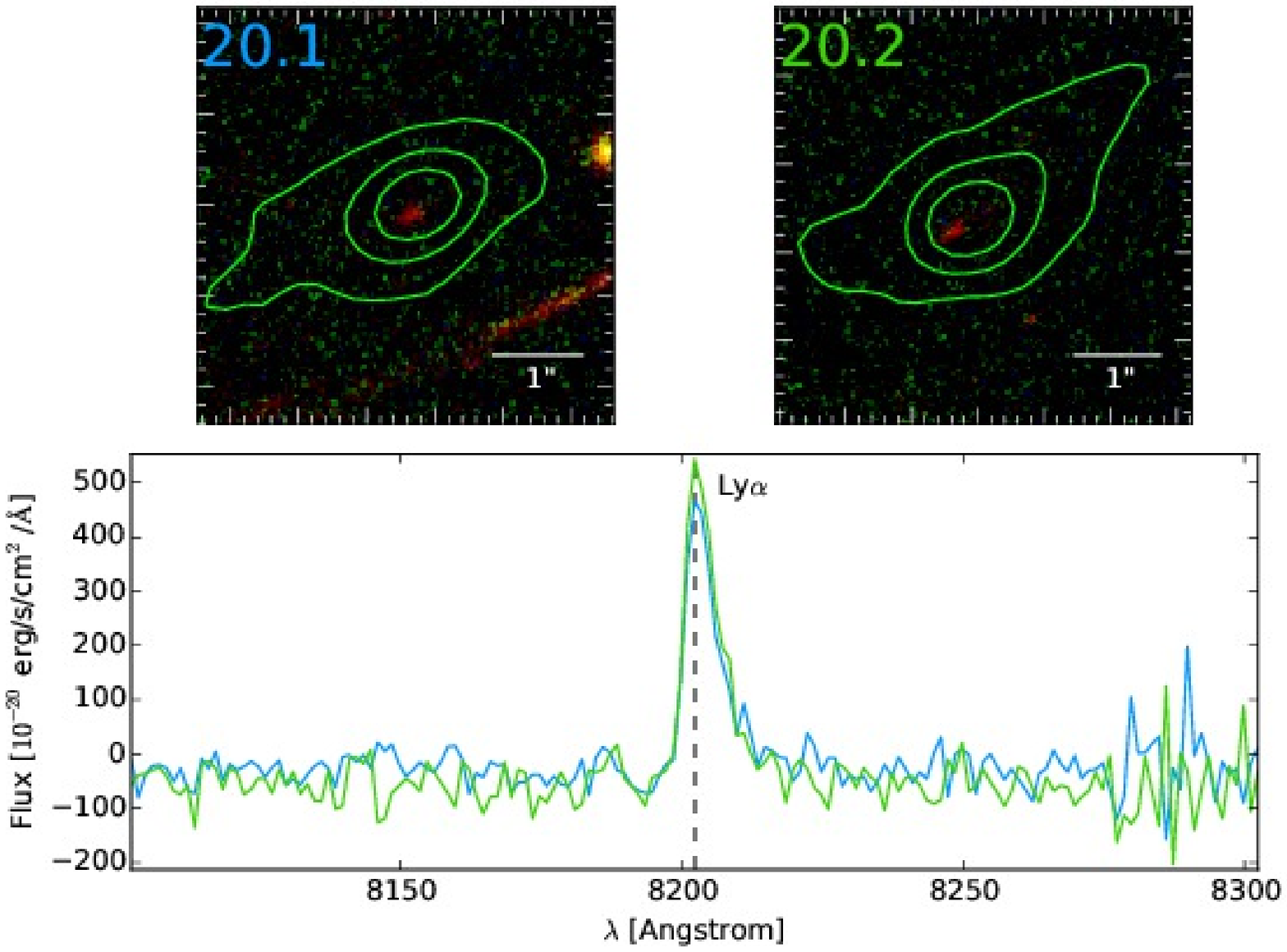}
    \includegraphics[width=0.5\textwidth]{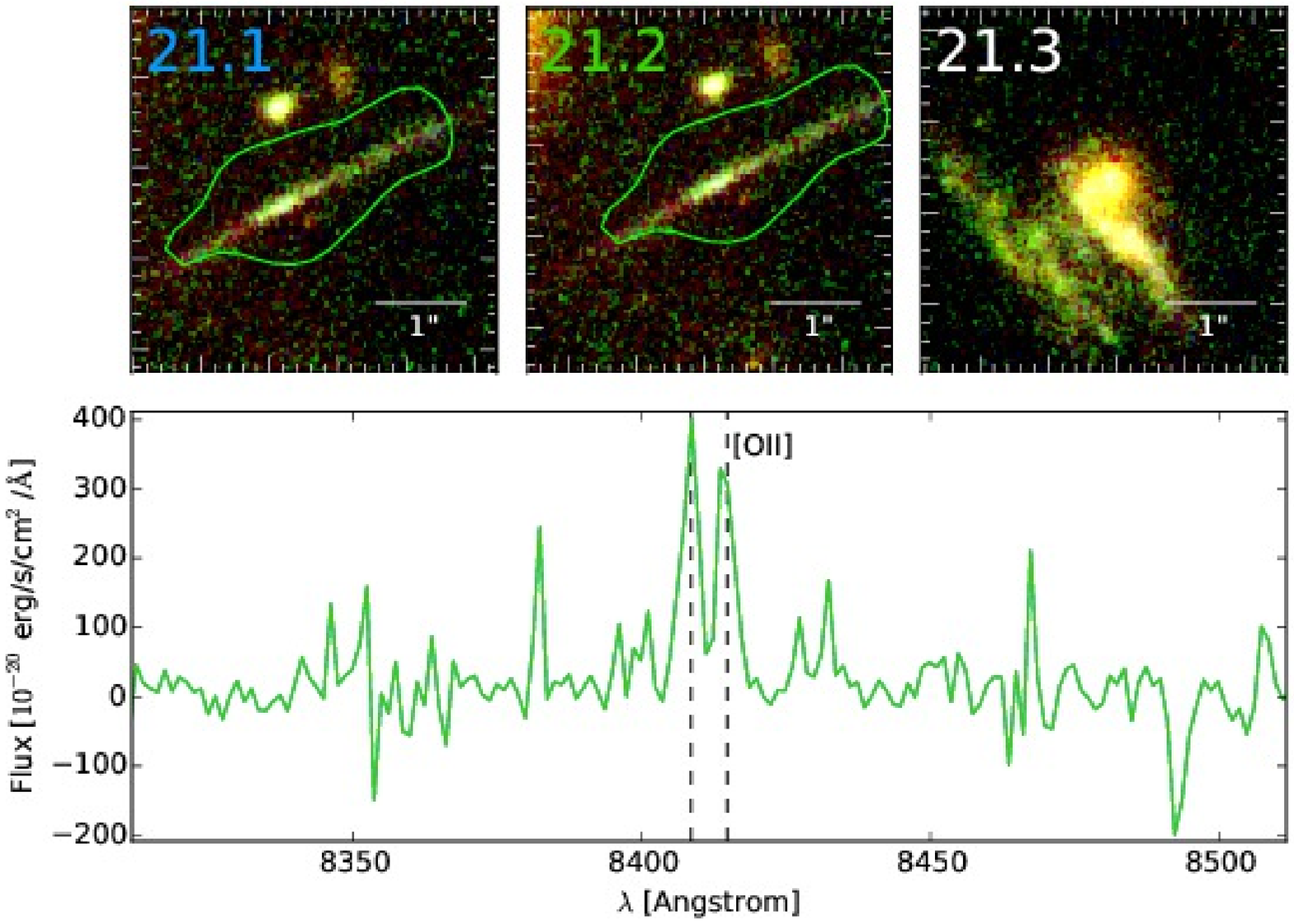}}
  \centerline{
    \includegraphics[width=0.5\textwidth]{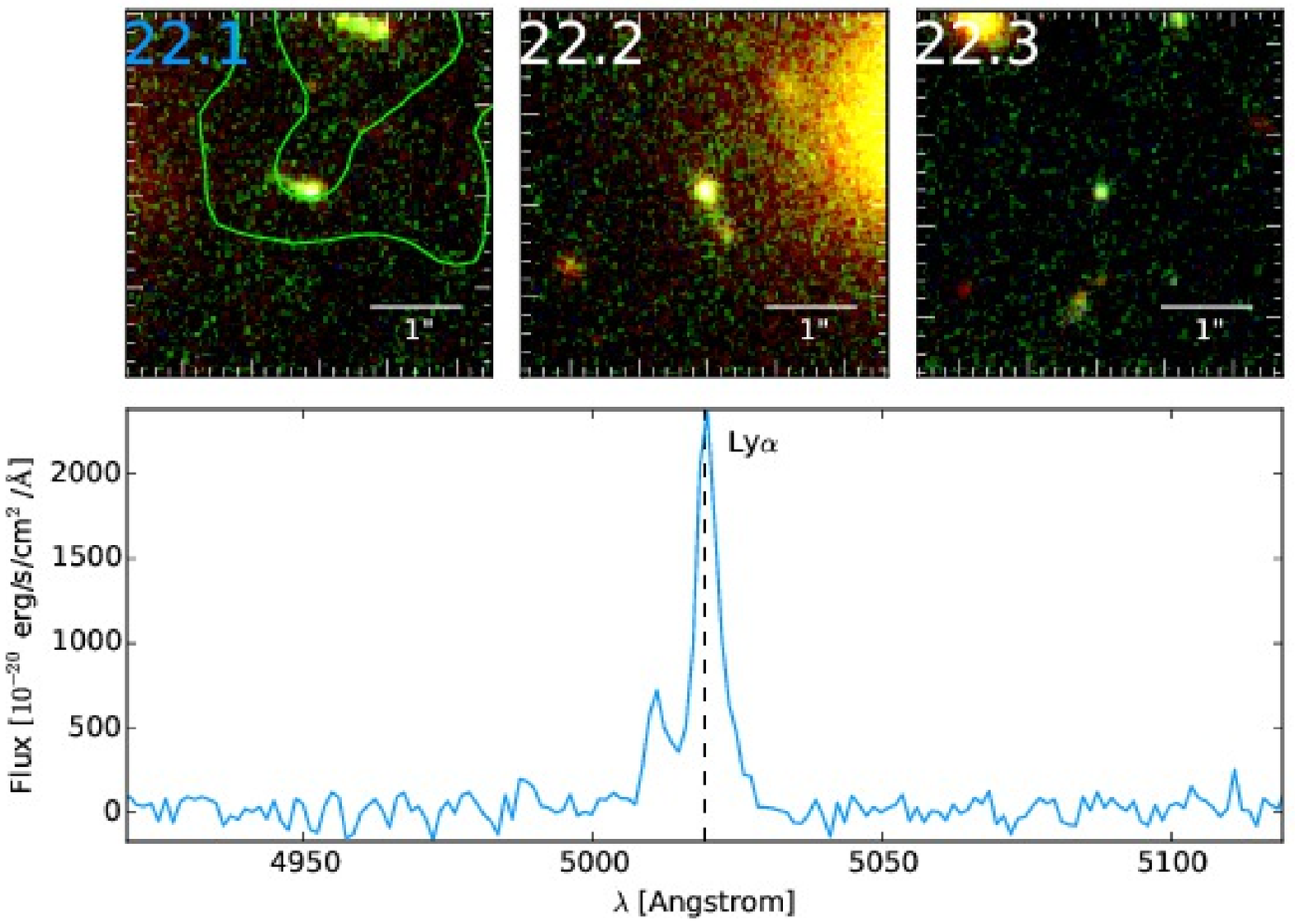}}
  \caption{(continued) The final three MUSE-identified systems.  We again
    see the overlap between the Lyman-$\alpha$ halos of Images 14.3
    and 22.1.}
  \label{fig:redshifts}
\end{figure*}

\subsection{Cluster Members}
\label{Cluster}
\begin{figure}
  \includegraphics[width=\columnwidth]{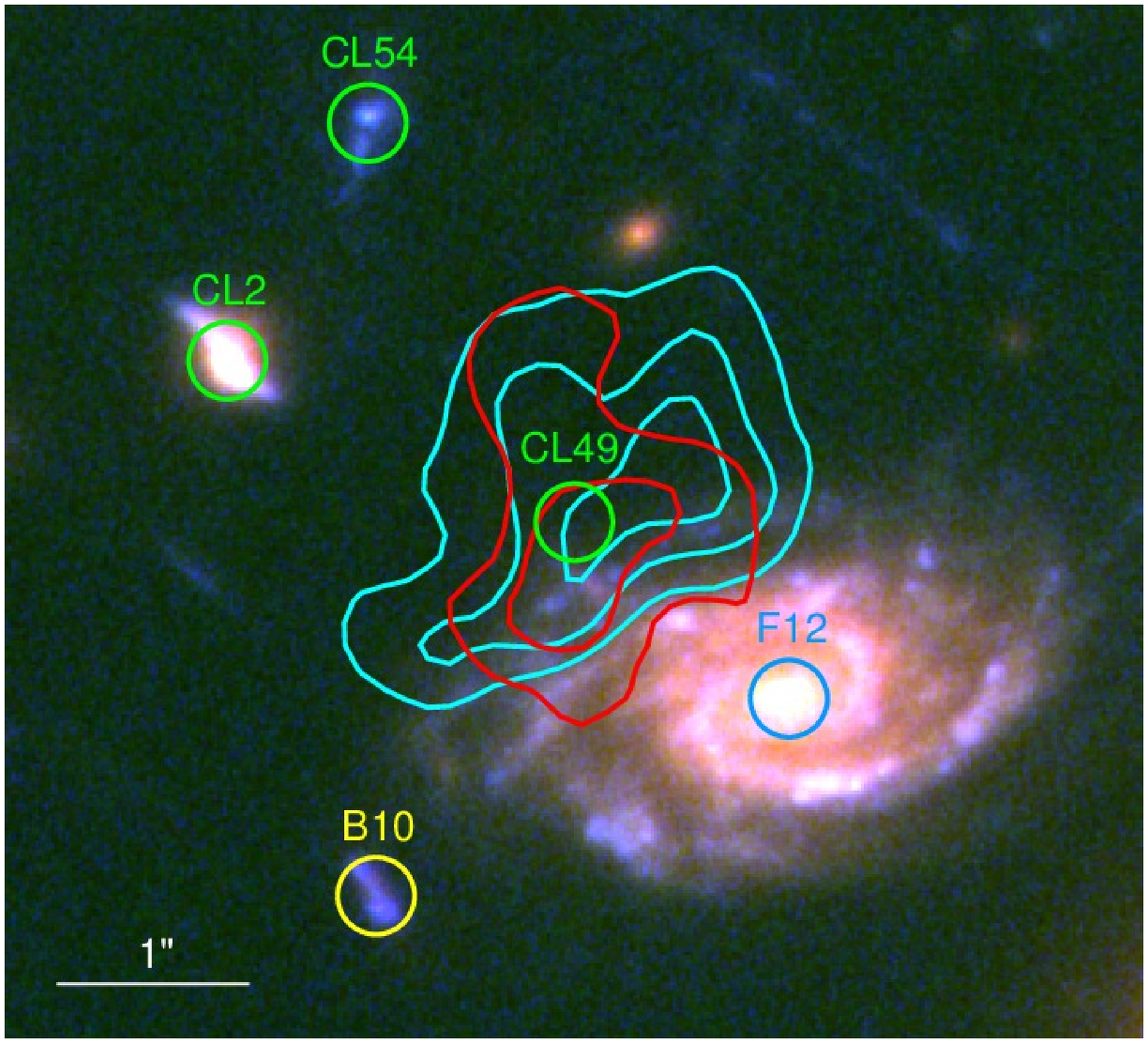}
  \caption{Cluster member CL49, an emission-line object ($z = 0.3843$)
    not seen in broadband HST imaging.  The estimated centroid of the
    object is shown as a green circle, while the cyan and red contours
    trace the 1-, 3-, and 5-$\sigma$ (per pixel) regions of [OII] and
    H$\alpha$ emission, respectively.  Curiously, the long axis of
    emission seems to have a divergent velocity field, as the flux at
    either end is slightly redshifted relative to the center.}
  \label{fig:MysteryObject}
\end{figure}

Cluster members make up the largest subgroup in the spectroscopic
catalog, with 56 confirmed redshifts.  To identify cluster members, we
select all galaxies that fall between $z = 0.35$ and $z = 0.4$, a
$\delta v = 15000$ km s$^{-1}$ cut in velocity space centered on the
A370 central redshift ($z = 0.375$).  In the overall redshift catalog,
these boundaries naturally separate the cluster overdensity from all
other objects, and a large fraction of galaxies in this range fall on
both the (F435W - F606W) and (F606W - F814W) red sequences.  While a
majority of objects are passive, elliptical galaxies showing only
absorption features, some do show strong [OII], [OIII], and/or
H$\alpha$ emission. Of these emission-line galaxies, CL49 (located
close to the bright, foreground spiral F12) is unique, as it is the
only detected cluster member in the MUSE field of view without an HST
counterpart (Figure \ref{fig:MysteryObject}.)  Strangely, the object
also appears to have a divergent velocity field: moving in either
direction along the long axis, the flux becomes more and more
redshifted relative to the center.  We do not have an explanation for
this behavior at this time, though it would be interesting to revisit
in future work.

Cluster galaxy positions are represented by green circles in the top
panel of Figure \ref{fig:Zdist}.  The redshifts themselves are
presented in Table \ref{tbl:ClusterZ}.

\begin{table*}
	\centering
	\caption{Cluster Members}
	\label{tbl:ClusterZ}
	\begin{tabular}{llllcccc}
		\hline
		 ID & RA & Dec & z & $m_{\rm F435W}$ & $m_{\rm F606W}$ & $m_{\rm F814W}$ & Type$^{~ \rm a}$\\
		\hline
 		 Cl1 & 39.975198 & -1.5879282 & 0.3582 & 23.98 & 22.31 & 21.55 & \ion{Ca}{II} H, K\\
		 Cl2 & 39.978460 & -1.5839292 & 0.3606 & 24.16 & 22.96 & 22.34 & H$\alpha$\\
		 Cl3 & 39.970141 & -1.5807560 & 0.3609 & 25.56 & 24.02 & 23.18 & \ion{Ca}{II} H, K\\
		 Cl4 & 39.967945 & -1.5844368 & 0.3624 & 24.47 & 22.56 & 21.59 & \ion{Ca}{II} H, K\\
		 Cl5 & 39.970898 & -1.5846121 & 0.3635 & 24.52 & 22.54 & 21.59 & \ion{Ca}{II} H, K\\
		 Cl6 & 39.967716 & -1.5866033 & 0.3639 & 22.83 & 20.89 & 19.86 & \ion{Ca}{II} H, K\\
		 Cl7 & 39.977458 & -1.5902598 & 0.3645 & 24.27 & 22.71 & 21.99 & \ion{Ca}{II} H, K\\
		 Cl8 & 39.974634 & -1.5833866 & 0.3648 & 26.09 & 24.55 & 23.72 & \ion{Ca}{II} H, K\\
		 Cl9 & 39.976794 & -1.5808331 & 0.3649 & 25.41 & 23.76 & 23.00 & \ion{Ca}{II} H, K\\
		Cl10 & 39.964373 & -1.5734012 & 0.3660 & 23.18 & 21.25 & 20.29 & \ion{Ca}{II} H, K\\
		Cl11 & 39.964968 & -1.5756005 & 0.3660 & 23.42 & 21.40 & 20.41 & \ion{Ca}{II} H, K\\
		Cl12 & 39.977446 & -1.5764519 & 0.3680 & 23.16 & 21.07 & 20.02 & \ion{Ca}{II} H, K\\
		Cl13 & 39.964290 & -1.5724542 & 0.3681 & 23.18 & 21.25 & 20.29 & \ion{Ca}{II} H, K\\
		Cl14 & 39.975885 & -1.5759172 & 0.3683 & 24.37 & 22.42 & 21.41 & \ion{Ca}{II} H, K\\
		Cl15 & 39.978375 & -1.5743572 & 0.3683 & 23.68 & 21.53 & 20.49 & \ion{Ca}{II} H, K\\
		Cl16 & 39.971815 & -1.5747844 & 0.3685 & 23.54 & 21.84 & 20.92 & \ion{Ca}{II} H, K\\
		Cl17 & 39.971961 & -1.5842487 & 0.3687 & 25.34 & 23.30 & 22.30 & \ion{Ca}{II} H, K\\
		Cl18 & 39.965402 & -1.5860164 & 0.3701 & 23.19 & 21.18 & 20.15 & \ion{Ca}{II} H, K\\
		Cl19 & 39.969599 & -1.5837975 & 0.3708 & 22.67 & 20.66 & 19.66 & H$\alpha$\\
		Cl20 & 39.970497 & -1.5748780 & 0.3708 & 24.06 & 22.01 & 21.01 & \ion{Ca}{II} H, K\\
		Cl21 & 39.963783 & -1.5810351 & 0.3710 & 21.76 & 19.82 & 18.81 & \ion{Ca}{II} H, K\\
                Cl22 & 39.971829 & -1.5860732 & 0.3711 & 24.65 & 22.77 & 21.87 & \ion{Ca}{II} H, K\\
		Cl23 & 39.965344 & -1.5760183 & 0.3716 & 23.62 & 21.57 & 20.54 & \ion{Ca}{II} H, K\\
		Cl24 & 39.977262 & -1.5819075 & 0.3718 & 23.37 & 21.18 & 20.07 & \ion{Ca}{II} H, K\\
                Cl25 & 39.972201 & -1.5803644 & 0.3727 & 24.83 & 23.12 & 22.24 & \ion{Ca}{II} H, K\\
		Cl26 & 39.973141 & -1.5768829 & 0.3728 & 23.88 & 21.94 & 20.95 & \ion{Ca}{II} H, K\\
		Cl27 & 39.973889 & -1.5764248 & 0.3728 & 25.00 & 23.21 & 22.25 & \ion{Ca}{II} H, K\\
                Cl28 & 39.973560 & -1.5743250 & 0.3728 & 25.05 & 23.17 & 22.19 & \ion{Ca}{II} H, K\\
		Cl29 & 39.971337 & -1.5822570 & 0.3731 & 22.00 & 19.74 & 18.66 & \ion{Ca}{II} H, K\\
		Cl30 & 39.974778 & -1.5798886 & 0.3738 & 25.44 & 24.60 & 24.26 & H$\alpha$\\
		Cl31 & 39.968403 & -1.5746894 & 0.3742 & 22.96 & 20.99 & 19.95 & \ion{Ca}{II} H, K\\
		Cl32 & 39.969132 & -1.5849674 & 0.3742 & 23.58 & 21.57 & 20.52 & \ion{Ca}{II} H, K\\
		Cl33 & 39.971118 & -1.5869043 & 0.3749 & 23.13 & 21.16 & 20.12 & \ion{Ca}{II} H, K\\
		Cl34 & 39.973823 & -1.5808796 & 0.3753 & 24.39 & 22.47 & 21.51 & \ion{Ca}{II} H, K\\
		Cl35 & 39.970597 & -1.5837833 & 0.3756 & 23.83 & 21.72 & 20.67 & \ion{Ca}{II} H, K\\
		Cl36 & 39.972478 & -1.5845667 & 0.3756 & 23.01 & 21.05 & 20.07 & \ion{Ca}{II} H, K\\
		Cl37 & 39.975144 & -1.5768716 & 0.3762 & 23.08 & 21.21 & 20.25 & \ion{Ca}{II} H, K\\
		Cl38 & 39.968075 & -1.5756317 & 0.3766 & 23.74 & 21.77 & 20.76 & \ion{Ca}{II} H, K\\
                Cl39 & 39.972565 & -1.5838926 & 0.3783 & 23.01 & 21.05 & 20.07 & \ion{Ca}{II} H, K\\
		Cl40 & 39.972358 & -1.5843052 & 0.3784 & 23.01 & 21.05 & 20.07 & \ion{Ca}{II} H, K\\
		Cl41 & 39.965603 & -1.5833191 & 0.3789 & 25.22 & 23.43 & 22.49 & \ion{Ca}{II} H, K\\
		Cl42 & 39.967642 & -1.5734009 & 0.3795 & 25.21 & 23.46 & 22.48 & \ion{Ca}{II} H, K\\
		Cl43 & 39.970183 & -1.5763142 & 0.3798 & 25.13 & 23.40 & 22.46 & \ion{Ca}{II} H, K\\
		Cl44 & 39.962843 & -1.5783866 & 0.3802 & 24.52 & 23.60 & 23.25 & H$\alpha$\\
		Cl45 & 39.971870 & -1.5797518 & 0.3807 & 22.85 & 21.71 & 21.07 & H$\alpha$\\
		Cl46 & 39.975789 & -1.5858086 & 0.3810 & 22.52 & 20.37 & 19.29 & \ion{Ca}{II} H, K\\
		Cl47 & 39.968845 & -1.5780882 & 0.3826 & 24.96 & 22.98 & 21.98 & \ion{Ca}{II} H, K\\
		Cl48 & 39.974799 & -1.5749206 & 0.3839 & 25.04 & 23.02 & 22.01 & \ion{Ca}{II} H, K\\
                Cl49 & 39.977573 & -1.5844846 & 0.3843 & 26.76 & 26.11 & 25.62 & [OII]\\
		Cl50 & 39.964628 & -1.5802868 & 0.3844 & 23.74 & 21.76 & 20.73 & \ion{Ca}{II} H, K\\
                Cl51 & 39.963103 & -1.5789205 & 0.3848 & 25.57 & 23.82 & 23.03 & \ion{Ca}{II} H, K\\
		Cl52 & 39.969230 & -1.5770399 & 0.3855 & 24.12 & 23.14 & 22.30 & \ion{Ca}{II} H, K\\
		Cl53 & 39.978147 & -1.5814394 & 0.3873 & 22.20 & 20.84 & 20.12 & H$\alpha$\\
		Cl54 & 39.978109 & -1.5833172 & 0.3873 & 24.16 & 22.96 & 22.34 & H$\alpha$\\
		Cl55 & 39.969088 & -1.5786944 & 0.3885 & 22.74 & 20.54 & 19.46 & \ion{Ca}{II} H, K\\
		Cl56 & 39.973107 & -1.5755088 & 0.3905 & 22.91 & 21.26 & 20.53 & \ion{Ca}{II} H, K\\
		\hline
	\end{tabular}
        \\
          $^{\rm a}$ Listed emission/absorption lines refer to the most prominent features seen in the galaxy's spectrum.
\end{table*}

\subsection{Other Objects}
\label{Interlopers}
In addition to cluster members and multiply-imaged systems, the
catalog also contains a number of foreground objects and singly-imaged
background objects at various redshifts.  We identify four stars
within the MUSE field of view, mostly located near the BCG.  Beyond
the Milky Way, we find 13 galaxies in the foreground of the cluster,
between $z = 0.2$ and $z = 0.35$.  Nearly all of these objects are
optically blue and have emission features, in particular a very
strong H-$\alpha$ line.  However, object F11 ($z = 0.3275$) shows
strong absorption line features, with only weak [OII] emission.
Behind the cluster, we find an additional 13 objects at redshifts
between $z = 0.41$ and $z = 1.5$.  Here again, these systems are
largely identified by strong [OII] and/or [OIII] emission lines,
however object B4 ($z = 0.4655$) is another absorption-line galaxy
showing weak [OII] emission.  Information on all singly-imaged objects
can be found in Table \ref{tbl:OtherZ}.

\begin{table}
	\centering
	\caption{Other Foreground and Background Objects}
	\label{tbl:OtherZ}
	\begin{tabular}{llllc}
		\hline
		 ID & RA & Dec & z & Type\\
		 \hline
		S1 & 39.972077 & -1.5805308 & 0.0000 & $\star$\\
		S2 & 39.970365 & -1.5859742 & 0.0000 & $\star$\\
		S3 & 39.973933 & -1.5873072 & 0.0000 & $\star$\\
		S4 & 39.964779 & -1.5801536 & 0.0000 & $\star$\\
		F1 & 39.978832 & -1.5754827 & 0.2067 & H$\alpha$\\
		F2 & 39.975921 & -1.5750762 & 0.2070 & [OIII]\\
		F3 & 39.978088 & -1.5746502 & 0.2181 & H$\alpha$\\
		F4 & 39.980006 & -1.5797900 & 0.2558 & H$\alpha$\\
		F5 & 39.972050 & -1.5744911 & 0.2559 & H$\alpha$\\
		F6 & 39.965942 & -1.5893178 & 0.3050 & H$\alpha$\\
		F7 & 39.966252 & -1.5838533 & 0.3247 & H$\alpha$\\
		F8 & 39.967436 & -1.5871793 & 0.3261 & H$\alpha$\\
		F9 & 39.978915 & -1.5750390 & 0.3263 & H$\alpha$\\
		F10 & 39.977545 & -1.5741827 & 0.3264 & H$\alpha$\\
		F11 & 39.977866 & -1.5779488 & 0.3275 & \ion{Ca}{II} H, K\\
		F12 & 39.977060 & -1.5847684 & 0.3461 & H$\alpha$\\
		F13 & 39.972236 & -1.5893030 & 0.3465 & H$\alpha$\\
		B1 & 39.973514 & -1.5800835 & 0.4104 & [OIII]\\
		B2 & 39.979117 & -1.5898644 & 0.4223 & [OIII]\\
		B3 & 39.976931 & -1.5909864 & 0.4225 & [OIII]\\
		B4 & 39.969400 & -1.5736444 & 0.4655 & \ion{Ca}{II} H, K\\
		B5 & 39.968908 & -1.5871056 & 0.5804 & [OIII]\\
		B6 & 39.969464 & -1.5895997 & 0.6037 & [OII]\\
		B7 & 39.967551 & -1.5861036 & 0.6801 & [OIII]\\
		B8 & 39.967227 & -1.5899252 & 0.8040 & [OIII]\\
		B9 & 39.963801 & -1.5882507 & 0.8049 & [OIII]\\
		B10 & 39.978094 & -1.5852708 & 1.0606 & [OII]\\
		B11 & 39.962676 & -1.5851813 & 1.0635 & [OII]\\
		B12 & 39.976042 & -1.5893204 & 1.3398 & [OII]\\
		B13 & 39.979281 & -1.5902451 & 1.4497 & [OII]\\                 
		\hline
	\end{tabular}
\end{table}

\section{Lens Modeling}
\label{Model}

To measure the mass distribution of A370, we follow a procedure used
in previous works \citep[e.g.,][]{ric14,jon14}: namely, we model the
system as a collection of mass clumps, including both large-scale dark
matter halos representing cluster potentials and smaller, galaxy-scale
halos representing individual galaxies.  Each halo, regardless of
size, is assumed to have a truncated Dual Pseudo-Isothermal Elliptical
mass distribution (dPIE; \citealt{eli07}).  The dPIE halo consists of
seven parameters: position ($\alpha$ and $\delta$), velocity
dispersion ($\sigma_0$), position angle ($\theta$), ellipticity
($\varepsilon$), and two scale radii ($r_{\rm core}$ and $r_{\rm
  cut}$) that modify the halo's mass slope.  $r_{\rm core}$ represents
the halo's inner core radius, inside of which the mass slope flattens
instead of increasing isothermally.  $r_{\rm cut}$ represents the
halo's cutoff radius, outside of which the mass slope drops more
steeply.

During the optimization process, we allow most cluster halo parameters
to freely vary within a given prior distribution, but we fix $r_{\rm
  cut}$ to a large value of 800 kpc (155\arcsec).  This is because the
current data do not extend far enough to meaningfully constrain its
value.  To limit the overall size of the parameter space, we place
additional restrictions on the galaxy-scale halos, adopting a
light-traces-mass approach that assumes correlation between the
observed baryons and their galaxy halos.  Galaxy positions,
ellipticities, and position angles are fixed to values measured from
the F814W HFF data, while $r_{\rm core}$, $r_{\rm cut}$, and
$\sigma_0$ are scaled based on the galaxy's luminosity ($L$), relative
to an $L^*$ galaxy. At the A370 cluster redshift, $m^*_{\rm F814W}$ =
18.31.  We fix $r^*_{\rm core}$ to be 0.15 kpc, while we allow
$r^*_{\rm cut}$ to vary with a uniform prior between 10 and 50 kpc.
The $r^*_{\rm cut}$ value is kept low to account for tidal stripping
effects by the cluster halos \citep{lim07b,hal07,nat09}.  Similarly,
we allow ($\sigma^*_0$) to vary, but we instead assume a Gaussian
prior distribution with mean $\mu_{\rm pdf}$ = 158 km s$^{-1}$ and
width $\sigma_{\rm pdf}$ = 27 km s$^{-1}$, following the \citet{ber03}
observations of early-type galaxies in high-density environments in
the Sloan Digital Sky Survey (SDSS).

Finally, we model three galaxy scale halos separately from the scaling
relation: the two BCGs and another bright galaxy (GAL1) lying close to
the System 2 giant arc.  The Southern BCG and GAL1 (objects Cl29 and
Cl32, respectively, in Figure \ref{fig:Zdist}) are close enough and
massive enough to significantly affect the shape of System 2.  The
Northern BCG mildly influences the orientation of the Northern cluster
potential.  Like the other galaxy potentials, we fix the position, PA,
and ellipticity values to match the F814W HFF data.  We also fix
$r_{\rm core}$, though we assume different values (0.14 kpc for the
BCGs, 0.06 kpc for GAL1) based on their magnitudes. $\sigma_0$ and
$r_{\rm cut}$ are allowed to vary freely, using their magnitude-scaled
values relative to $L^*$ as a starting point.

\begin{figure}
  \includegraphics[width=\columnwidth]{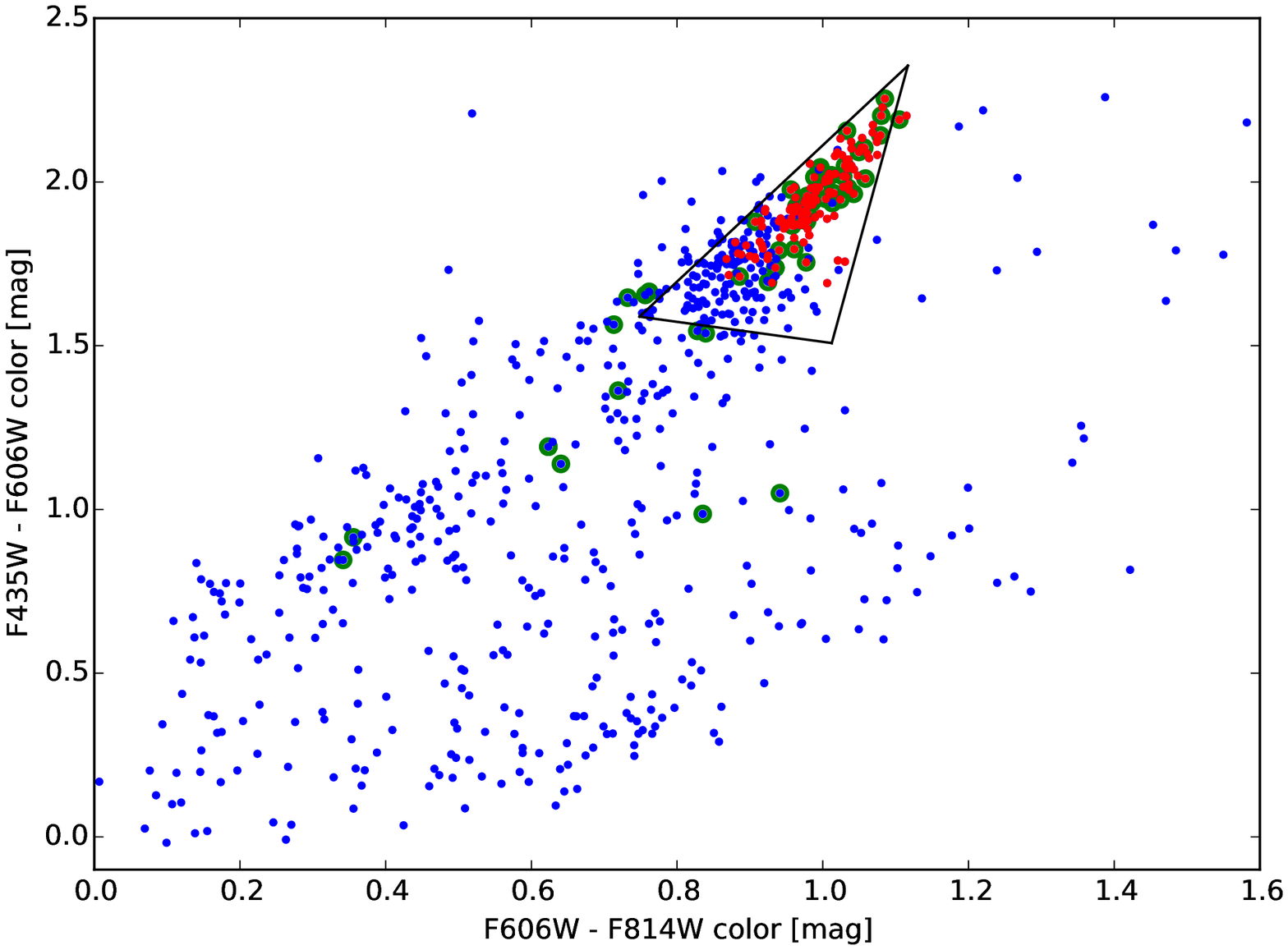}
    \caption{Color-color criterion used to select galaxies for mass
      modeling.  ``Good'' objects -- those falling simultaneously
      within 3-$\sigma$ of the F435W-F606W vs $m_{\rm F814W}$ and
      F606W-F814W vs $m_{\rm F814W}$ cluster red-sequences -- are
      mainly located within the black triangle region, though an
      additional faint-end magnitude cutoff ($m_{\rm F814W} < 22.6$)
      eliminates several candidates.  The final set of selected
      galaxies is shown in red.  Additionally, spectroscopically
      confirmed cluster members are highlighted by green rings, and
      are included in the final model regardless of color.}
    \label{fig:ColorColor}
\end{figure}

Individual galaxies are selected through a color cut, using the three
optical HFF bands.  In particular, we construct the F435W-F606W and
F606W-F814W color-magnitude diagrams and look for evidence of a
red-sequence.  We then combine the two plots into a color-color
diagram and select galaxies which fall into the tight cluster locus
(Figure \ref{fig:ColorColor}).  To measure galaxy photometry, we run
\texttt{SExtractor} \citep{ber96} in dual-image mode, using the F814W
band data as the detection template.  In the region covered by the
MUSE footprint, we further refine the selection process by including
all spectroscopically-confirmed cluster members and rejecting all non
cluster members regardless of their colors.  Additionally, we limit
the selection to bright galaxies ($M_{\rm F814W} < 22.6$), since
beyond this cutoff -- roughly equivalent to $\sim$ 0.02 $L_*$ --
galaxies will have a negligible effect on the mass budget.  We also
exclude galaxies at distances beyond $R = 70$\arcsec\ from the cluster
center ($\alpha =$ $2^{\rm h}$ $39^{\rm m}$ $59\fs9$, $\delta =$
$-1\degr$ $34\arcmin$ $36\farcs 5$), corresponding to $R > 360$ kpc in
physical space, for similar reasons.

\begin{figure}
  \includegraphics[width=\columnwidth]{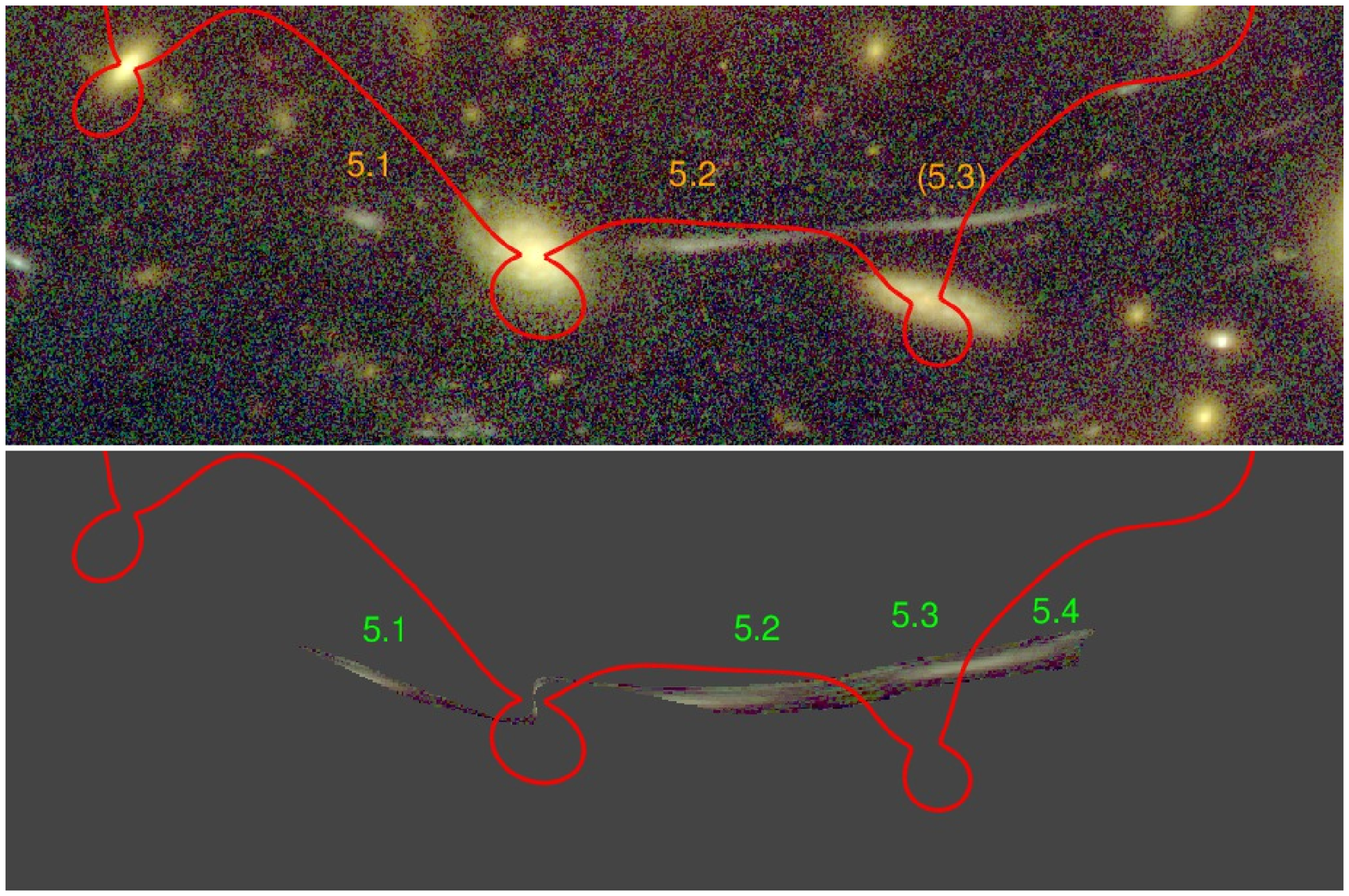}
  \caption{Our new interpretation of System 5. {\bf Top:} an ACS RGB
    image of System 5. {\bf Bottom:} A simulated image, using the lens
    model.  In this simulation, a pixelized image of the galaxy is
    sent to the source plane and then back to the image plane to be
    compared with observations. The critical line at $z = 1.2775$ (red
    line in both panels) is affected by the background $z = 0.6$
    spiral galaxy (B6; see table \ref{tbl:OtherZ}) on the right and
    crosses the rightmost image previously known as 5.3. We interpret
    5.3 as a merging pair of two additional images mirroring only part
    of the source.  This is seen convincingly in the simulation, which
    reproduce the relative surface brightnesses of the observations
    well.}
  \label{fig:Sys5Sim}
\end{figure}

We constrain the mass model with the positions of known
multiply-imaged systems, however, we first revise the interpretation
of System 5, previously described as a 3-image system (5.1, 5.2,
5.3). Image 5.3 is actually located close to a background galaxy at $z
= 0.6$ (object B6 in Table \ref{tbl:OtherZ}) which perturbs the lens
model, and we add it as an additional galaxy-scale potential.  As
illustrated by a simulation of the lensing effect on Images 5.1 and
5.2, this small perturbation convincingly produces a merging pair of
two additional images previously identified as 5.3 (Figure
\ref{fig:Sys5Sim}.)  Therefore, we update our constraint list to
include the new Image 5.3 and Image 5.4 in the model.  While we use
nearly all of the 22 current systems (Table \ref{tbl:Multi-Images}) we
exclude System 10, a faint radial arc without a secure redshift that
also lacks bright counterimages.  Because even small changes in the
mass model produce wildly different predictions for radial systems,
the lack of counterimages makes the system fully degenerate with the
parameters of the Northern cluster halo.

While some systems have only a single MUSE detection (Systems 16, 17,
18, 19, and 22), these objects are often bright in broadband
photometry, and we are able to identify their counterparts in the HFF
data directly.  In fact, many lens systems with MUSE spectroscopy have
at least one counterimage that falls outside of the current footprint,
since the GTO data only cover a fraction of the A370 field of view.
Though we are unable to confirm these objects spectroscopically, we
can identify them with the help of the lens model itself.
Specifically, we start with known (spectroscopically confirmed)
members of a given lens system and use the model to predict the
location of counterimages.  From this rough guess, we look for
candidates in the HFF data that match the system members (i.e.,
galaxies with similar colors, morphology, and shape), which we take to
be the counterimages.  We then optimize a new version of the model,
including these other constraints.  We can compare the models with and
without the added constraints, and if the new results stay within an
acceptable error limit: an average model rms displacement of
1\arcsec\ or lower, we keep the counterimages and start the process
again.  In this way, we are able to iteratively refine and improve the
model until it converges on the best solution.

In the best A370 model, we identify 16 counterimages this way: Images
7.4, 7.5, 14.4, 14.5, 15.4, 15.5, 16.3, 17.2, 17.3, 18.2, 18.3, 19.2,
19.3, 21.3, 22.2, and 22.3.  With the exception of the System 7
counterimages, they are all associated with newly-identified MUSE
objects, again highlighting the power of MUSE as a redshift engine.
While the model also predicts the existence of Images 3.4 and 20.3,
there are no obvious candidates in the available HFF data, so we do
not include them as constraints.  The remaining systems (8, 11, 12,
and 13) fall outside of the MUSE footprint and lack secure redshift
information.  Thus, we use the positions of these systems as
constraints but leave their redshift values as free parameters to be
fit by the model.  In some cases, the best-fit model predicts
additional counterimages for these objects (Images 8.3, 11.3, and
13.3), but we again find no obvious counterparts in the HFF data, so
we do not use them as model constraints.  This is because predicted
positions without redshift have significantly larger uncertainties.
However, this will be revisited in future work with a larger MUSE
mosaic.  After adding and removing objects as described above, we are
left with 21 unique systems with a total of 66 individual image
constraints.

To optimize the model we feed all parameters and constraints into the
\texttt{LENSTOOL}\footnote{\url{https://projets.lam.fr/projects/lenstool/wiki}}
\citep{kne96,jul07,jul09} software program, which probes the full
parameter space with a Bayesian Markov Chain Monte Carlo (MCMC)
sampler.  To evaluate a given model, \texttt{LENSTOOL} reconstructs
the parametrized mass distribution and transforms the constraint
coordinates from the image (observed) plane to the source
(undeflected) plane.  In the source plane, the program calculates the
barycenter of each constraining system, then transforms these results
back to the lens plane, creating a set of predicted images that can be
compared to observation.  By minimizing the rms displacement between
prediction and observation, \texttt{LENSTOOL} is thus able to
objectively determine the ``best'' model for a given set of input
parameters (e.g., \citealt{kne96,lim07a}).

Following \citet{ric14}, our initial model includes two massive
cluster potentials (each centered on a BCG) and the cluster galaxies
selected from the color-color cut outlined above.  However, after
several MCMC iterations with this model, we find that no combination
of parameter values sufficiently recreates the observed image
positions (i.e., with an average rms error $<$ 1\arcsec.)  In
particular, we are unable to simultaneously model the positions of the
systems in the North and the new MUSE-identified images near the
center.  This is most noticeable in System 12, where typical model
displacements can be as as large as 6\arcsec.

To help correct this disagreement, we first add an elongated, bar-like
mass potential (DM3) as a bridge between the two cluster halos (Table
\ref{tbl:ModParams}).  This flattens the mass distribution in the
region between the two BCGs, and improves the fit of the central,
radial image constraints of Systems 7, 14, and 15.  As a consequence
of adding the bar, the Northern cluster potential (DM2) becomes more
elongated and its centroid moves further away from the BCG.  This can
be explained as a model degeneracy between the positions of DM2 and
DM3, since there is a significant overlap between their mass
distributions.  In fact, the barycenter of the two components falls
slightly to the South of the BCG itself.

\begin{figure}
  \includegraphics[width=\columnwidth]{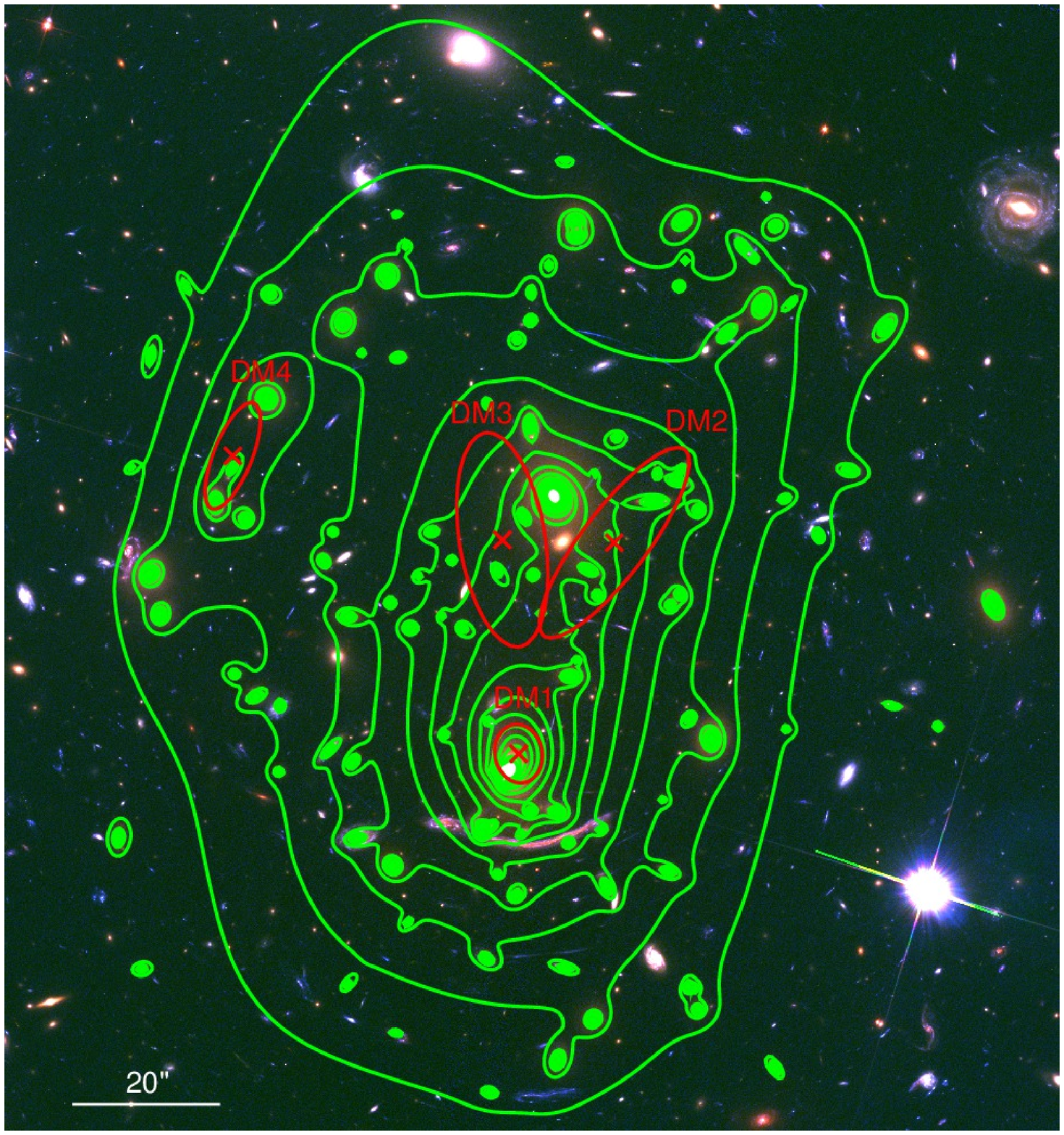}
  \caption{Dark matter surface mass density contours (green) measured
    from the lens model.  Contours are shown in steps of
    $3\times10^{8}$ $M_{\odot}$ kpc$^{-2}$, starting from $10^9$
    $M_{\odot}$ kpc$^{-2}$.  Highlighted in red are the centroids of
    the four cluster-scale dark matter halos.  Red ellipses are used
    to give a rough estimate of the relative shape and strength of
    each component, but we stress that the actual mass contours are
    much larger and flatter.}
    \label{fig:massCont}
\end{figure}

Even after adding the bar, however, we are only able to reduce the
model rms to 1.17\arcsec, with the Northern systems still poorly
fit. Therefore, we add an additional large-scale cluster potential to
the model (DM4), centered near a group of bright galaxies in the
northeastern corner of the cluster (see Figure \ref{fig:massCont}).
This location is chosen both because of its proximity to Image 12.3
(the image constraint with the highest rms error), and because it is
part of a ``crown'' of galaxies sitting above the northern BCG.  Since
all of the galaxies in the crown are bright and have similar,
cluster-like colors and several are spectroscopically confirmed to be
cluster members \citep{mel88} it is possible this collection
represents another, unaccounted-for mass component of A370.

As with the other cluster potentials, we allow the dPIE parameters of
the crown component to vary freely within a set of priors, though we
assume an initial configuration that is more elliptical (like the
bar), to better account for the stretched nature of the crown.
Including this additional mass clump further improves the fit,
reducing the average rms displacement from 1.17\arcsec\ to
0.94\arcsec, below the benchmark level of rms = 1\arcsec.

To see if even more mass structures are needed to describe A370, we
also test a 5-clump mass model, placing an additional mass clump on
the Western side of the A370 crown.  However, this does not
significantly improve the fit (rms = 0.88\arcsec) and actually lowers
the Bayesian evidence, possibly suggesting an overfit to the data (see
Section \ref{Clump3}.)  Thus, given the success of the 4-clump model
in reproducing the image constraint positions, we treat its optimized
parameter set as our ``best-fit''description of the A370 mass
distribution.  Optimized parameters for all models can be seen in
Table \ref{tbl:ModParams}, and the final model mass reconstruction is
shown in Figure \ref{fig:massCont}.

\begin{table*}
  \centering
  \caption{Candidate Lens Models and Best-Fit Parameters}
  \label{tbl:ModParams}
  \begin{tabular}{lrrrrrrrr}
    \hline
    Model Name & Component & $\Delta\alpha^{\rm ~a}$& $\Delta\delta^{\rm ~a}$ & $\varepsilon^{\rm ~b}$ & $\theta$ & $r_{\rm core}$ & $r_{\rm cut}$ & $\sigma_0$\\
    (Fit Statistics) &  & (\arcsec) & (\arcsec) & & ($\deg$) & (kpc) & (kpc) & (km s$^{-1}$)\\
    \hline
    2-Clump & DM1 & $ -0.69^{+  0.04}_{ -0.06}$ & $  1.83^{+  0.32}_{ -0.10}$ & $ 0.50^{+ 0.01}_{-0.01}$ & $-73.1^{+  0.5}_{ -0.3}$ & $ 51.4^{+  0.4}_{ -1.0}$ & $[800.0]^{\rm ~c}$ & $786^{+2}_{-5}$ \\[4pt]
    rms = 2.81\arcsec & DM2 & $  4.74^{+  0.12}_{ -0.15}$ & $ 36.64^{+  0.11}_{ -0.14}$ & $ 0.47^{+ 0.01}_{-0.01}$ & $-104.9^{+  0.3}_{ -0.3}$ & $170.4^{+  2.8}_{ -1.3}$ & $[800.0]$ & $1345^{+7}_{-4}$ \\[4pt]
    $\chi^2/\nu$ = 31.97 & BCG1 & $[ -0.01]$ & $[  0.02]$ & $[0.30]$ & $[-81.9]$ & $[  0.14]$ & $ 46.5^{+  1.0}_{ -1.0}$ & $205^{+8}_{-2}$ \\[4pt]
    $\log$($\mathcal{L}$) = -1652.47 & BCG2 & $[  5.90]$ & $[ 37.24]$ & $[0.20]$ & $[-63.9]$ & $[  0.14]$ & $ 48.3^{+  4.2}_{ -5.6}$ & $177^{+34}_{-7}$ \\[4pt]
    $\log$($\mathcal{E}$) = -1817.07 & CL32 & $[  7.92]$ & $[ -9.76]$ & $[0.26]$ & $[ 25.7]$ & $[  0.06]$ & $  7.4^{+  1.0}_{ -1.2}$ & $90^{+6}_{-1}$ \\[4pt]
    BIC = 3404.75 & L$^{*}$ galaxy & -- & -- & -- & -- & $[0.15]$ & $ 44.4^{+  0.8}_{ -2.5}$ & $163^{+2}_{-2}$\\[4pt]
    \hline
    3-Clump & DM1 & $  1.57^{+  0.13}_{ -0.08}$ & $  3.05^{+  0.01}_{ -0.09}$ & $ 0.36^{+ 0.02}_{-0.03}$ & $-78.7^{+  1.3}_{ -3.3}$ & $ 22.3^{+  0.6}_{ -1.0}$ & $[800.0]$ & $564^{+4}_{-3}$ \\[4pt]
    rms = 1.17\arcsec & DM2 & $ 19.58^{+  1.68}_{ -0.50}$ & $ 32.65^{+  1.12}_{ -2.75}$ & $ 0.90^{+ 0.03}_{-0.05}$ & $-130.6^{+  1.1}_{ -0.9}$ & $149.8^{+  6.1}_{ -7.5}$ & $[800.0]$ & $665^{+20}_{-33}$ \\[4pt]
    $\chi^2/\nu$ = 6.00 & DM3 & $ -3.29^{+  0.39}_{ -0.35}$ & $ 37.02^{+  0.97}_{ -0.98}$ & $ 0.54^{+ 0.02}_{-0.01}$ & $ 91.4^{+  1.2}_{ -0.8}$ & $210.4^{+  3.0}_{ -4.6}$ & $[800.0]$ & $1376^{+26}_{-9}$ \\[4pt]
    $\log$($\mathcal{L}$) = -209.75 & BCG1 & $[ -0.01]$ & $[  0.02]$ & $[0.30]$ & $[-81.9]$ & $[  0.14]$ & $ 61.3^{+  5.2}_{ -2.5}$ & $177^{+11}_{-5}$ \\[4pt]
    $\log$($\mathcal{E}$) = -303.19 & BCG2 & $[  5.90]$ & $[ 37.24]$ & $[0.20]$ & $[-63.9]$ & $[  0.14]$ & $ 63.2^{+  1.9}_{ -6.4}$ & $317^{+1}_{-15}$ \\[4pt]
    BIC = 554.49 & CL32 & $[  7.92]$ & $[ -9.76]$ & $[0.26]$ & $[ 25.7]$ & $[  0.06]$ & $ 15.3^{+  2.5}_{ -4.5}$ & $74^{+3}_{-4}$ \\[4pt]
    & $L^{*}$ Galaxy & -- & -- & -- & -- & $[0.15]$ & $ 45.7^{+  2.1}_{ -2.3}$ & $230^{+9}_{-1}$\\[4pt]
    \hline
    4-Clump & DM1 & $  1.10^{+  0.19}_{ -0.17}$ & $  2.40^{+  0.18}_{ -0.10}$ & $ 0.22^{+ 0.03}_{-0.02}$ & $-79.3^{+  4.6}_{ -5.2}$ & $ 20.4^{+  0.6}_{ -1.0}$ & $[800.0]$ & $536^{+4}_{-6}$ \\[4pt]
    rms = 0.94\arcsec & DM2 & $ 14.04^{+  0.73}_{ -0.48}$ & $ 30.95^{+  0.92}_{ -0.81}$ & $ 0.85^{+ 0.01}_{-0.01}$ & $-127.0^{+  0.3}_{ -0.3}$ & $148.4^{+  3.9}_{ -2.3}$ & $[800.0]$ & $917^{+10}_{-12}$ \\[4pt]
    $\chi^2/\nu$ = 4.33 & DM3 & $ -1.11^{+  0.28}_{ -0.36}$ & $ 31.34^{+  1.29}_{ -1.03}$ & $ 0.73^{+ 0.01}_{-0.01}$ & $ 99.2^{+  0.5}_{ -0.5}$ & $179.9^{+  5.5}_{ -3.7}$ & $[800.0]$ & $1159^{+9}_{-11}$ \\[4pt]
    $\log$($\mathcal{L}$) = -146.48 & DM4 & $-37.58^{+  1.15}_{ -1.02}$ & $ 42.69^{+  1.22}_{ -1.20}$ & $ 0.79^{+ 0.02}_{-0.02}$ & $ 68.2^{+  1.5}_{ -1.1}$ & $ 55.3^{+  6.2}_{ -3.1}$ & $[800.0]$ & $530^{+8}_{-11}$ \\[4pt]
    $\log$($\mathcal{E}$) = -201.90 & BCG1 & $[ -0.01]$ & $[  0.02]$ & $[0.30]$ & $[-81.9]$ & $[  0.14]$ & $ 38.1^{+  4.0}_{ -3.5}$ & $185^{+17}_{-6}$ \\[4pt]
    BIC = 455.35 & BCG2 & $[  5.90]$ & $[ 37.24]$ & $[0.20]$ & $[-63.9]$ & $[  0.14]$ & $ 52.4^{+  3.9}_{ -2.5}$ & $316^{+12}_{-30}$ \\[4pt]
    & CL32 & $[  7.92]$ & $[ -9.76]$ & $[0.26]$ & $[ 25.7]$ & $[  0.06]$ & $ 28.8^{+  6.4}_{ -5.2}$ & $88^{+5}_{-5}$ \\[4pt]
    & $L^{*}$ Galaxy & -- & -- & -- & -- & $[0.15]$ & $ 38.2^{+  1.0}_{ -2.6}$ & $197^{+6}_{-8}$\\[4pt]
    \hline
     5-Clump & DM1 & $  1.44^{+  0.18}_{ -0.17}$ & $  2.35^{+  0.33}_{ -0.11}$ & $ 0.17^{+ 0.03}_{-0.03}$ & $-73.5^{+  6.8}_{ -4.3}$ & $ 24.4^{+  1.5}_{ -1.1}$ & $[800.0]$ & $561^{+4}_{-3}$ \\[4pt]
     rms = 0.88\arcsec & DM2 & $  9.96^{+  0.32}_{ -1.19}$ & $ 27.54^{+  0.57}_{ -1.55}$ & $ 0.83^{+ 0.02}_{-0.03}$ & $-124.3^{+  0.8}_{ -0.9}$ & $129.8^{+  4.5}_{ -5.2}$ & $[800.0]$ & $857^{+21}_{-34}$ \\[4pt]
     $\chi^2/\nu$ = 4.29 & DM3 & $ -2.29^{+  0.41}_{ -0.88}$ & $ 36.56^{+  2.28}_{ -0.88}$ & $ 0.76^{+ 0.02}_{-0.03}$ & $100.0^{+  0.8}_{ -0.8}$ & $182.4^{+  3.4}_{ -8.5}$ & $[800.0]$ & $1134^{+22}_{-23}$ \\[4pt]
     $\log$($\mathcal{L}$) = -133.64 & DM4 & $-38.41^{+  1.66}_{ -1.24}$ & $ 44.48^{+  1.43}_{ -1.70}$ & $ 0.80^{+ 0.02}_{-0.02}$ & $ 73.5^{+  4.2}_{ -2.3}$ & $ 51.4^{+  8.8}_{ -4.3}$ & $[800.0]$ & $531^{+21}_{-18}$ \\[4pt]
     $\log$($\mathcal{E}$) = -202.67 & DM5 & $ 44.27^{+  0.11}_{ -1.71}$ & $ 53.79^{+  8.08}_{ -0.28}$ & $ 0.75^{+ 0.04}_{-0.14}$ & $-35.1^{+  2.7}_{ -5.3}$ & $ 36.9^{+  4.5}_{ -8.7}$ & $[800.0]$ & $436^{+52}_{-57}$ \\[4pt]
     BIC = 456.27 & BCG1 & $[ -0.01]$ & $[  0.02]$ & $[0.30]$ & $[-81.9]$ & $[  0.14]$ & $ 40.6^{+  5.3}_{ -4.9}$ & $200^{+11}_{-14}$ \\[4pt]
     & BCG2 & $[  5.90]$ & $[ 37.24]$ & $[0.20]$ & $[-63.9]$ & $[  0.14]$ & $ 51.0^{+  5.0}_{ -2.6}$ & $280^{+17}_{-12}$ \\[4pt]
     & CL32 & $[  7.92]$ & $[ -9.76]$ & $[0.26]$ & $[ 25.7]$ & $[  0.06]$ & $  5.9^{+  4.6}_{ -5.7}$ & $101^{+7}_{-6}$ \\[4pt]
   &  $L^{*}$ Galaxy & -- & -- & -- & -- & $[0.15]$ & $ 46.1^{+  2.9}_{ -1.6}$ & $199^{+9}_{-10}$\\[4pt]
    \hline
  \end{tabular}
  \medskip\\
  $^{\rm a}$ $\Delta\alpha$ and $\Delta\delta$ are measured relative to the reference coordinate point: ($\alpha$ = 39.97134, $\delta$ = -1.5822597)~~~~~~~~~~~~~~~~~~~~~~~~~~~~~~~\\[1pt]
  $^{\rm b}$ Ellipticity ($\varepsilon$) is defined to be $(a^2-b^2) / (a^2+b^2)$, where $a$ and $b$ are the semi-major and semi-minor axes of the ellipse~~~~~~\\[1pt]
   $^{\rm c}$ Quantities in brackets are fixed parameters~~~~~~~~~~~~~~~~~~~~~~~~~~~~~~~~~~~~~~~~~~~~~~~~~~~~~~~~~~~~~~~~~~~~~~~~~~~~~~~~~~~~~~~~~~~~~~~~~~~~~~~~~
\end{table*}

\section{Discussion}
\label{Discus}

\subsection{The Case for Additional Mass Clumps}
\label{Clump3}

While adding two new large-scale mass clumps to the model
significantly improves the fit, it is important to ask the question
``Is this additional mass truly necessary?''  In other words, are
these new mass clumps real features supported by the data, or is it
simply a case of overfitting in a sparsely sampled region of the
cluster?  To answer this, we look at both model and (perhaps more
importantly) physical evidence.

From a purely statistical point of view, the Bayesian evidence value
($\mathcal{E}$) can discriminate between models, taking into account
and in some cases penalizing additional terms and model complexities,
while simultaneously evaluating goodness of fit.  Models with larger
evidence terms are preferred over those with smaller terms, providing
an objective criterion for comparison (see e.g., \citealt{lim10}).
For the A370 models (Table \ref{tbl:ModParams}), we find that the
evidence of the 3-clump model ($\log(\mathcal{E})$ = -303.19) is
significantly larger than that of the 2-clump model
($\log(\mathcal{E})$ = -1817.07), showing that the data strongly
prefer a mass distribution that includes a flatter, central ``bar''
component.  The evidence term of the 4-clump model is larger still
($\log(\mathcal{E})$ = -201.90) -- though the relative improvement
over the 3-clump model is not as great -- again suggesting a real need
for the crown clump.  Finally, although the more complex 5-clump model
has a marginally improved fit over the 4-clump model (rms =
0.88\arcsec), its evidence value actually decreases ($\log
\mathcal{E}$ = -202.67) suggesting that the fifth mass clump is
unnecessary and that we are beginning to overfit the system.
Additional tests, placing the fifth clump at several other locations
throughout the crown, yield similar (or worse) results.

As a complementary check on overfitting, we also calculate the
  Bayesian information criterion (BIC) term for each model:
\begin{equation}
{\rm BIC} = -2 \times \log(\mathcal{L}) + k \times \log(n),
\end{equation}
where $\mathcal{L}$ is the model likelihood, $k$ is the number of
model free parameters, and $n$ is the number of model constraints.
Lower BIC values are favored over higher values, and for our models,
we again see that the 4-clump case (BIC = 455.35) is preferred over
all others (we present all values in Table \ref{tbl:ModParams}),
bolstering the claim that the bar and the crown are real, necessary
features.

\begin{figure*}
\centerline{
    \includegraphics[width=\columnwidth]{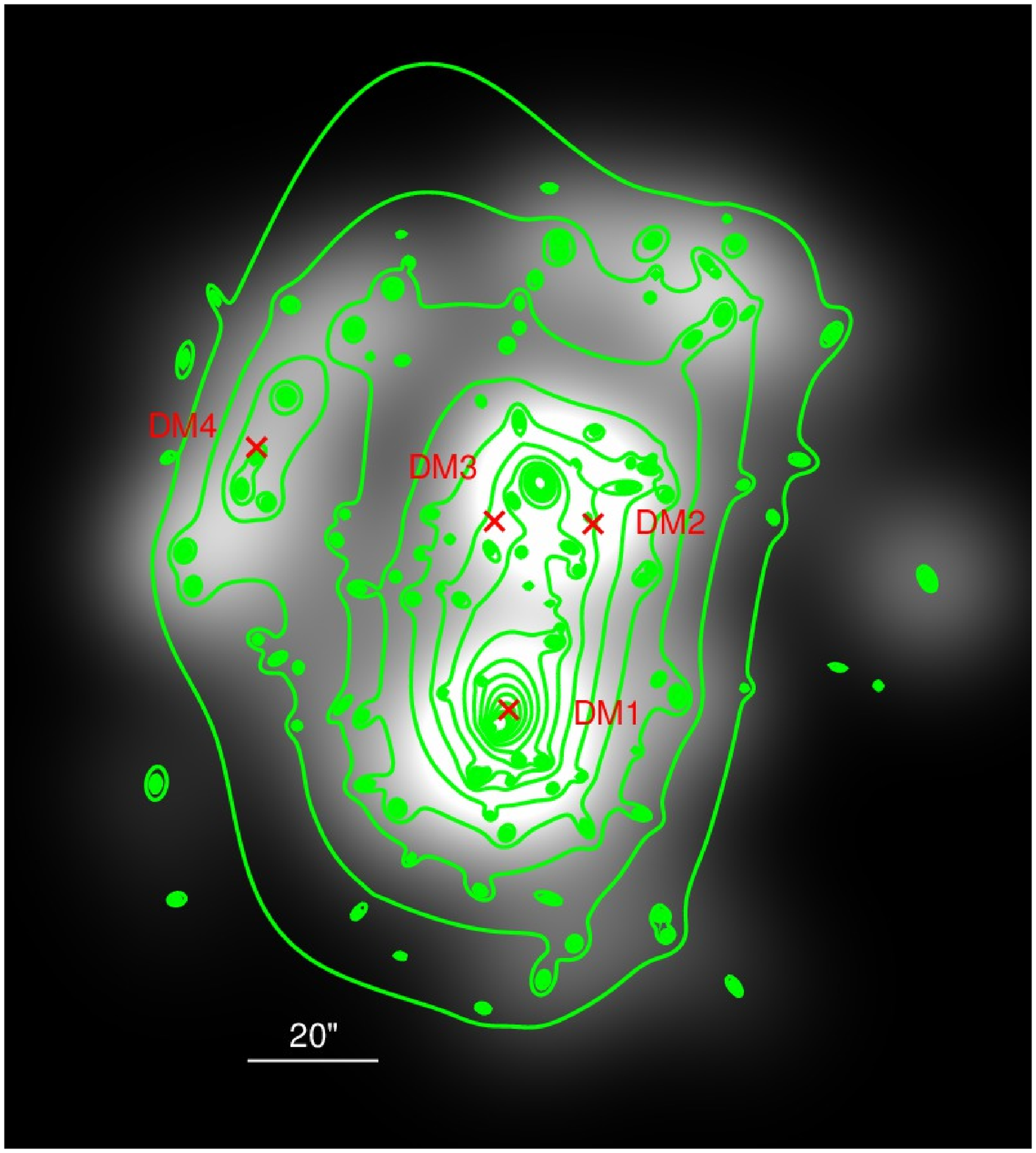}
    \includegraphics[width=\columnwidth]{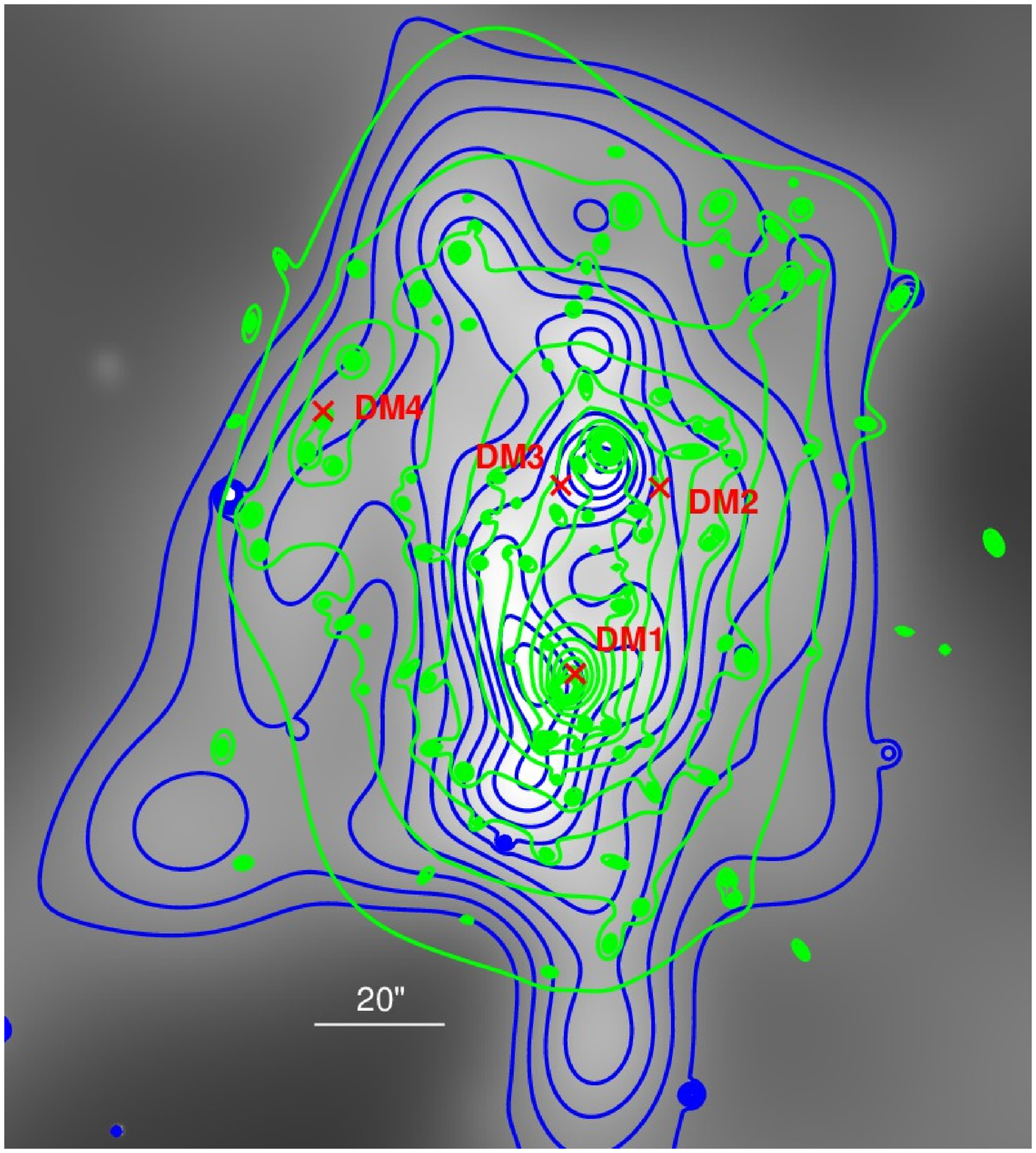}}
\caption{Comparisons between model mass clumps and two physical
  parameter maps.  {\bf Left:} The smoothed cluster light map.  We see
  an agreement between the mass clump positions and the locations of
  the brightest cluster light.  We also find a good agreement between
  the orientation of the light in the ``crown'' and the fourth mass
  clump (DM4).  {\bf Right:} The X-ray gas map.  We again see an
  agreement between the map and the large-scale clump positions, along
  with an agreement in shape and orientation of the X-ray gas contours
  (blue lines) and model mass contours (green lines.) In both cases
  the mass contours are the same as those presented in Figure
  \ref{fig:massCont}. Taken together these two maps suggest that the
  bar (DM3) and the crown (DM4) are real features of the cluster, and
  not a simple case of overfitting the data.}
\label{fig:Compare}
\end{figure*}

On the physical side of the evidence argument, we first turn to the
cluster light distribution.  As previously mentioned, the presence of
the Northern crown of galaxies already suggests an additional mass
distribution not captured by the two-clump model.  To test this
theory, we first isolate the cluster light in the F814W band (the
filter where cluster members are brightest), keeping only the objects
selected by the cluster-member color-color cut (Section \ref{Model})
and masking the rest.  We then smooth the remaining light with a
Gaussian kernel ($\sigma_{\rm smooth} = 10$\arcsec), providing a
cleaner view of the light distribution.  Comparing this distribution
to the 4-clump mass model (Figure \ref{fig:Compare}, left), we find
good agreement between the brightest points of the light map and the
positions of the large-scale clumps.  As expected, the Southern
cluster halo sits near the southern BCG, the combined Northern cluster
halo and bar surround the Northern BCG, and the fourth clump sits on
the Eastern side of the crown.  Even more promisingly, the orientation
of the fourth mass clump is aligned with the distribution of the crown
light, even though it is allowed to vary freely in the model.
Assuming then that, at least to some degree, the presence of light
traces the presence of mass, the agreement between the light map and
mass model provides another argument in favor of the 4-clump model.

Finally, we also look at the A370 X-ray gas profile, as the presence
of hot X-ray gas is often used as a tracer of deep mass potentials.
For this, we turn to publicly available, deep Chandra data: an 88 ks
image of A370, observed using the using the Advanced CCD Imaging
Spectrometer S-array (ACIS-S) camera (ID: 08700025, PI: G. Garmire).
After smoothing the X-ray map using the \texttt{ASMOOTH} algorithm
\citep{ebe06}, we compare the smoothed X-ray map to the positions of
the 4-clump mass model components (Figure \ref{fig:Compare}, right).
While the results here are not as definitive as in the cluster-light
case, we do still see a moderate agreement between the map and the
clump positions.  Generally, the X-ray contours are also slightly
extended in the vicinity of the crown clump, again hinting at the
presence of an additional potential well.

Focusing more closely on the X-ray contours, we can see that they
follow a distinct box-like pattern, which is especially noticeable in
the outskirts of the cluster.  Interestingly, the mass contours in the
4-clump model have a similar shape, largely driven by the overlap
between the Northern cluster halo and the central bar.  Given the
rough agreement between the X-ray and mass contours, it is possible
that the best-fit positions of the Northern mass clumps are more than
a simple degeneracy, but instead driven by physical parameters.

Thus, taken as a whole, the combination of physical and statistical
evidence indicate that the additional mass clumps in the cluster
center and crown region are necessary components of the mass model.
This suggests an even more complex mass distribution than previously
thought, motivating future studies in this area.

\subsection{Comparisons with Other Models}

Thanks to the discovery of several new lensing constraints, our
picture of the A370 mass distribution is beginning to evolve.  While
the addition of new mass clumps is the largest change, it is not the
only difference.  To get a better understanding of these changes and
what they mean physically, we can compare our model to previous
results.  This is especially easy with the models presented by
\citet{jon14} and \citet{ric14}, since they were both also generated
with \texttt{LENSTOOL}, using a similar set of parameters.  On the
whole, while the total mass of A370 remains largely unchanged between
models, where the mass enclosed within a 500 kpc circular radius is
$\sim 8\times10^{14} ~M_{\odot}$ in each case, the distribution of
mass between various components does not.  In particular, we find that
the cluster halo associated with the Southern BCG (DM1) is much
rounder and more compact than either of the previous models,
flattening the mass profile in this region.  This is predominantly due
to the number of radial, 5-image systems seen near the Southern
cluster (Systems 7, 14, and 15), as a more elliptical mass
distribution would break the symmetry and destroy the radial arcs.
Conversely, the Northern cluster halo (DM2) is more elliptical and
tilted at a steeper angle away from North, though this is largely
caused by an interaction with the central bar component.  Both cluster
potentials are also considerably less massive in our model, which is
another consequence of the bar.  Finally, we note that our model
favors a significantly larger $L^*$ galaxy velocity dispersion,
placing more emphasis on localized substructure contributions to the
total mass profile.  A list of specific parameter differences can be
seen in Table \ref{tbl:ModCompare2}.

\begin{table*}
  \centering
  \caption{Model Parameter Comparisons}
    \label{tbl:ModCompare2}
  \begin{tabular}{ccccccc}
    \hline
    Parameter & Model$^a$ & $\sigma_0$ & $\varepsilon$ & $\theta$ & $r_{\rm core}$ & $r_{\rm cut}$\\
     &        &  (km s$^{-1}$)     &             & (degree) &  (kpc)        & (kpc)\\
    \hline
    DM1      & J14 & 969$^{+100}_{-46}$  & 0.47$^{+0.02}_{-0.03}$ &   80.8$^{+0.99}_{-0.74}$ &  88.2$^{+8.7}_{-5.7}$  & [1500]$^b$\\[2pt]
            & R14 & 833$^{+58}_{-6}$  & 0.59$^{+0.04}_{-0.04}$ & -106.0$^{+2.8}_{-3.3}$ &  64.0$^{+8.0}_{-5.0}$  & [1000]\\[2pt]
            & L16 & $536^{+4}_{-6}$ & $ 0.22^{+ 0.03}_{-0.02}$ & $-79.3^{+  4.6}_{ -5.2}$ & $ 20.4^{+  0.6}_{ -1.0}$ & $[800.0]$\\[2pt]
    DM2      & J14 & 1040$^{+45}_{-120}$ & 0.09$^{+0.02}_{-0.06}$ &   89.4$^{+10.0}_{-3.9}$ &  94.7$^{+3.7}_{-15.0}$  & [1500]\\[2pt]
            & R14 & 1128$^{+37}_{-51}$ & 0.38$^{+0.04}_{-0.05}$ &  -89.6$^{+2.8}_{-2.4}$ & 155.0$^{+9.0}_{-12.0}$  & [1000]\\[2pt]
            & L16 & $917^{+10}_{-12}$ & $ 0.85^{+ 0.01}_{-0.01}$ & $-127.0^{+  0.3}_{ -0.3}$ & $148.4^{+  3.9}_{ -2.3}$ & $[800.0]$\\[2pt]
    $L^*$ Galaxy  & J14 & [120] & --- & --- & [0.15] & [120]\\[2pt]
            & R14 & 116$^{+16}_{-8}$   & --- & --- & [0.15] & 61.0$^{+21.0}_{-5.0}$\\[2pt]
    & L16 & $197^{+6}_{-8}$ & --- & --- & $[0.15]$ & $ 38.2^{+  1.0}_{ -2.6}$\\[2pt]
    
    \hline
  \end{tabular}
  \medskip\\
  $^a$ Model Key -- J14: \citet{jon14}, R14: \citet{ric14}, L16: This work\\
  $^b$ Quantities in brackets are fixed parameters~~~~~~~~~~~~~~~~~~~~~~~~~~~~~~~~~~~~~~~~~~~~~~~~~~~~~
  \end{table*}

\begin{figure*}
\centerline{
  \includegraphics[width=\columnwidth]{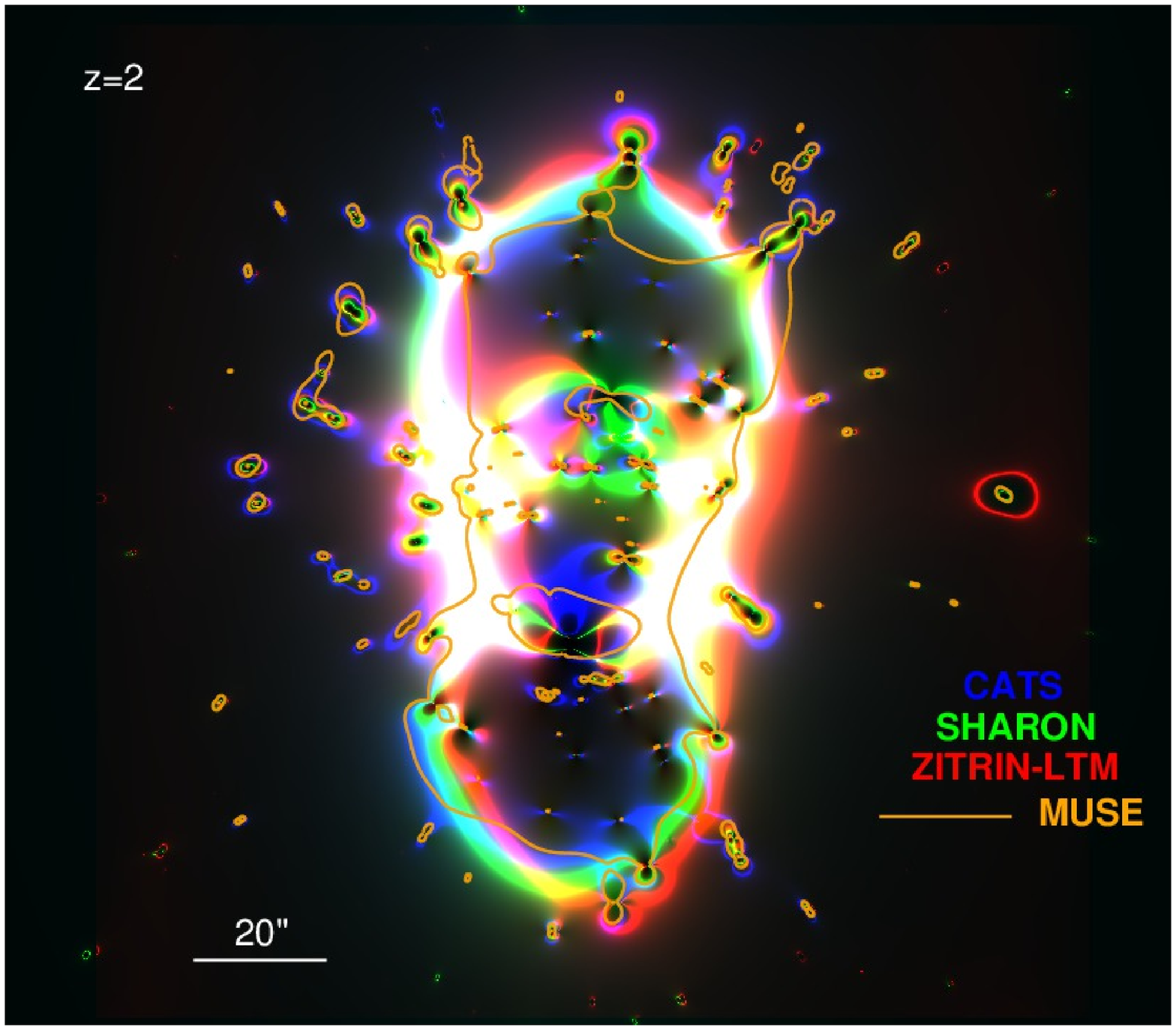}
  \includegraphics[width=\columnwidth]{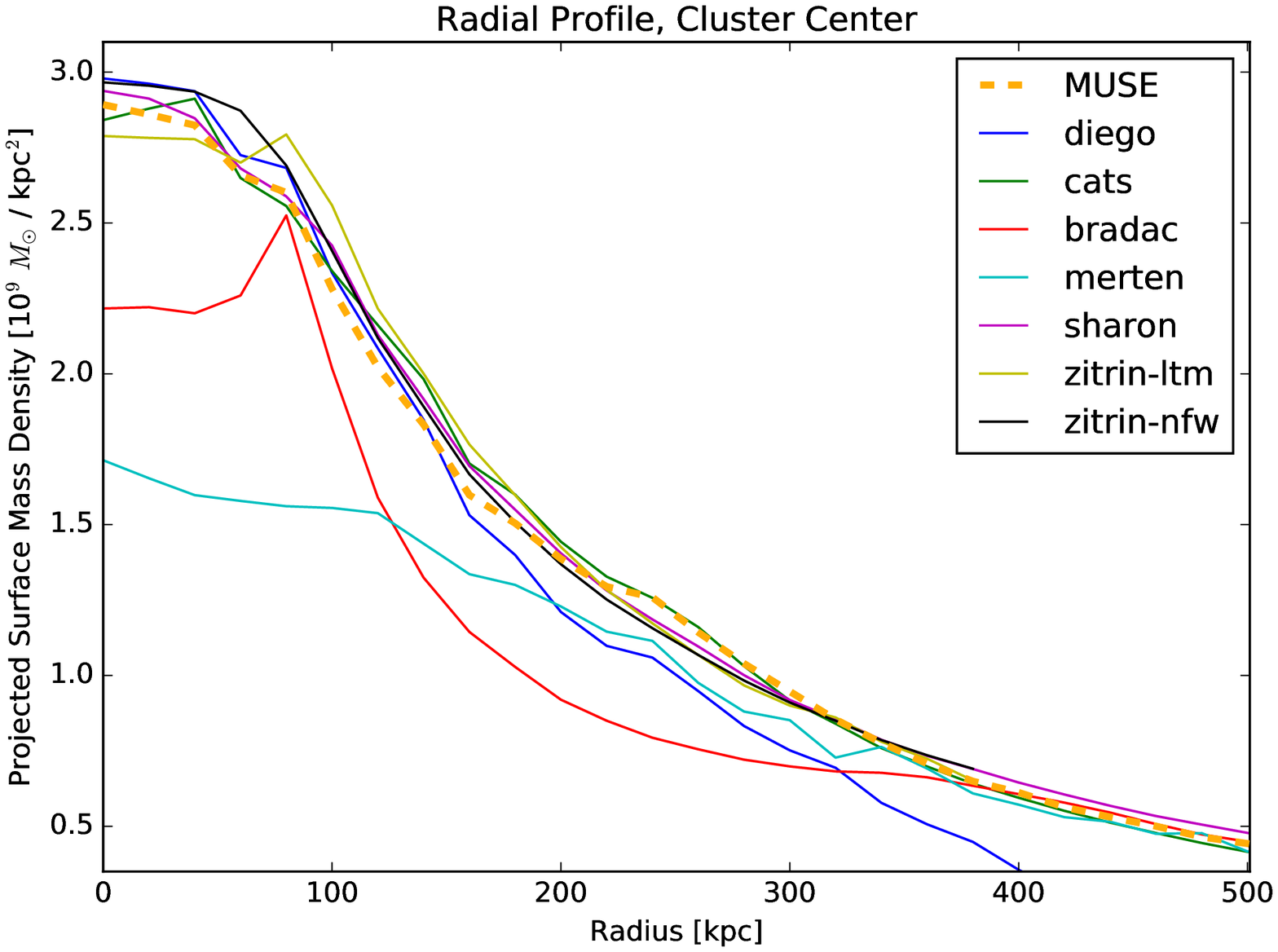}}
\caption{Comparisons between our best-fit mass model and previous HFF
  models.  {\bf Left:} Magnification maps of the CATS (blue), Sharon
  (green), and Zitrin-LTM (red) models, compared to the $z = 2$
  critical curve of our model (orange line).  We see good agreement
  between our model and the others, but our model has a larger radial
  critical region (internal orange line) driven by the discovery of
  several radial lens systems.  Additionally, our critical line
  extends further north than the other models, due to a larger, more
  elliptical Northern cluster potential (Table \ref{tbl:ModParams}.)
  {\bf Right:} Radial surface mass density profile, using several HFF
  public models.  Our model (black line) has a lower value at small
  radii, due to the radial caustic region.  At $\sim$ 250 kpc, we see
  a distinct bump, which represents the additional mass clump in the
  crown.}
\label{fig:Compare2}
\end{figure*}

In addition to specific model parameters, we can also compare the
total mass properties of our model.  For this exercise, we turn to the
public models constructed for the HFF lens modeling initiative.  In
the left panel of Figure \ref{fig:Compare2}, we show the $z = 2$
critical curve for our best-fit model, compared to the magnification
maps of three different HFF models: ``CATS'', from the Clusters As
TelescopeS team, (Co-PIs: J.P. Kneib and P. Natarajan), ``Sharon v.2''
(PI, K. Sharon), and ``Zitrin-LTM'' (PI, A. Zitrin).  We select these
models because they are constructed using only strong-lensing
constraints, and because the resolution of their magnification maps at
the core of the cluster is sufficiently high enough to compare their
critical curves.  In general, the shape of our curve largely traces
the high-magnification regions in the other models, suggesting a broad
agreement between our work and earlier studies.  However, we do note
that our curve has a boxier shape, due to the interaction between the
Northern cluster halo and the central bar and to the presence of the
crown clump.  Furthermore, the radial critical region (traced out by
the internal orange line) is much larger in our model.  This is again
a result of the rounder, flatter mass distribution predicted by our
model, and also highlights the increased area containing radial
lensing constraints such as Systems 7, 14, and 15.

These features can also be seen in the right-hand panel of Figure
\ref{fig:Compare2}, where we show the radial surface mass density
profile for several models (including \citealt{die16}), starting from
a point roughly centered between the two BCGs ($\alpha =$ $2^{\rm h}$
$39^{\rm m}$$52\fs937$, $\delta =$ $-1\degr$ $34\arcmin$ $37\farcs
003$).  At small radii ($<$ 100 kpc), our mass profile is lower than
several others, since this region covers the flatter core responsible
for the extended radial critical curve.  On the other hand, our model
profile is slightly higher than the others between 200 and 300 kpc,
due to the crown mass clump.

\begin{figure}
  \includegraphics[width=\columnwidth]{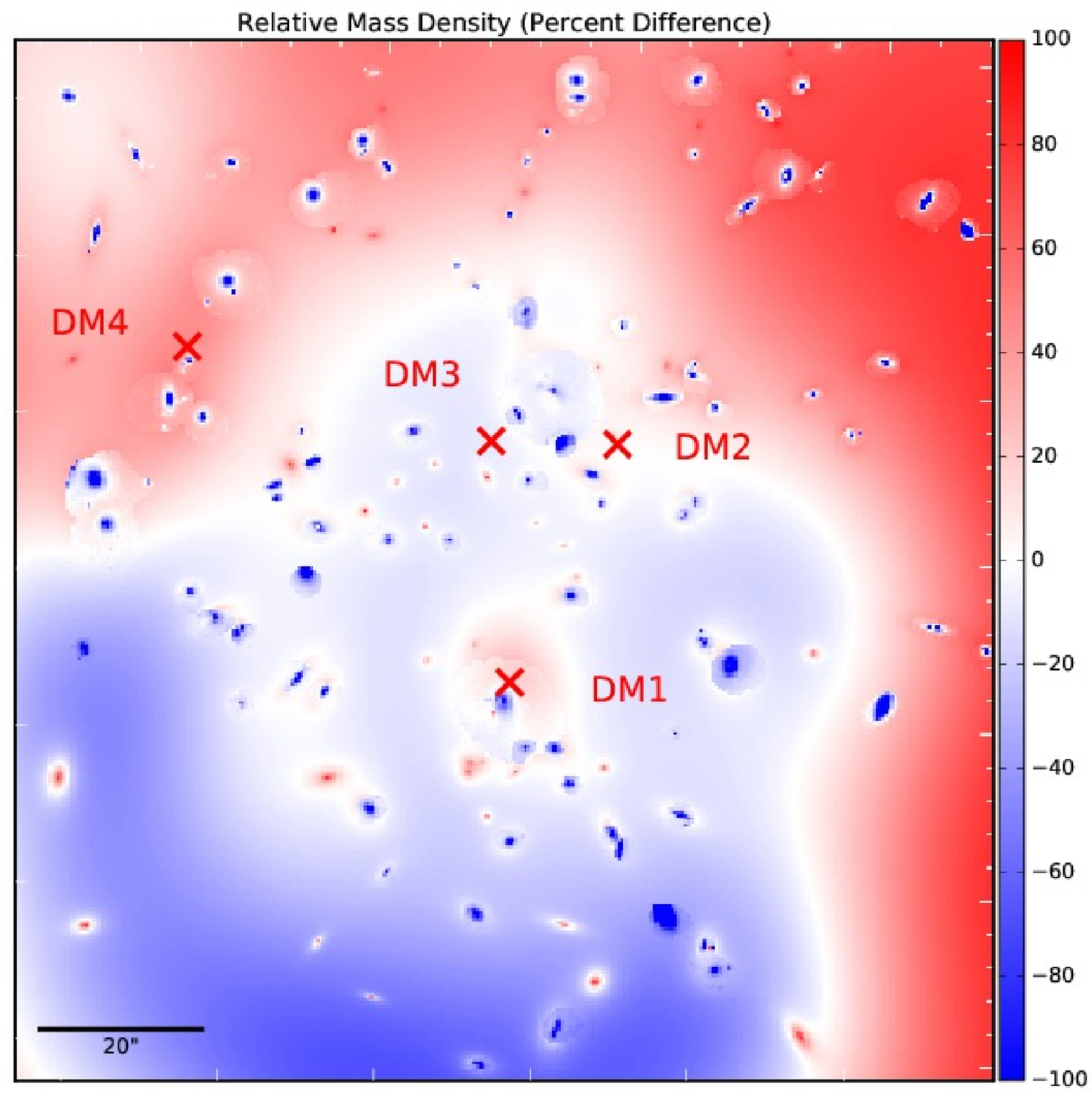}
  \caption{Residual mass density map between our best parametric model
    and the free-form model presented in \citet{die16}. In this plot,
    red represents an overdensity in the parametric model, while blue
    represents an overdensity in the free-form case.  Although there
    are differences in the spatial distributions of the models, the
    typical variation is less than ten percent.  We note that the
    largest deviations appear in the cluster outskirts, where the
    absolute mass density is small.  As a result, the total mass is
    roughly the same in both cases.}
    \label{fig:2dMap}
\end{figure}

Of course, important spatial information is lost when the mass map is
radially averaged, so it is important to look for differences in the
full 2D mass distribution, as well.  For this test, we compare our
4-clump model to the free-form model presented in \citet{die16}, which
is also generated from HFF data, but constructed in a completely
independent way, reducing potential sources of unknown bias.  To make
the comparison, we subtract the free-form model's mass density map
from our own map, measuring relative differences on a pixel-by-pixel
basis.  The results can be seen in Figure \ref{fig:2dMap}.  Overall,
we do see differences between the two approaches: The \citet{die16}
model is generally more concentrated and slightly rounder.  Our
parametric approach shows a noticeable overdensity in the vicinity of
the crown, due to the crown mass clump (DM4) in the East, and the
extension of the Northern cluster clump and the central bar (DM2 and
DM3, respectively) in the Northwest.  The parametric model also has a
larger fraction of total mass in the outskirts of the cluster, which
is also seen in the radial profile presented in Figure
\ref{fig:Compare2}.  While model differences can be large at large
radii -- the parametric model is rougly twice as dense as the free-form
model at the edges of the map -- we note that the absolute mass density
values at these distances are very small ($\sim 10^8 M_{\odot}$
kpc$^{-2}$) and contribute a negligible amount to the total mass
budget.  Conversely, the relative differences near the cluster center
are typically less than 10 percent, resulting in an overall good
agreement between the two models.

\begin{figure*}
\includegraphics[width=2\columnwidth]{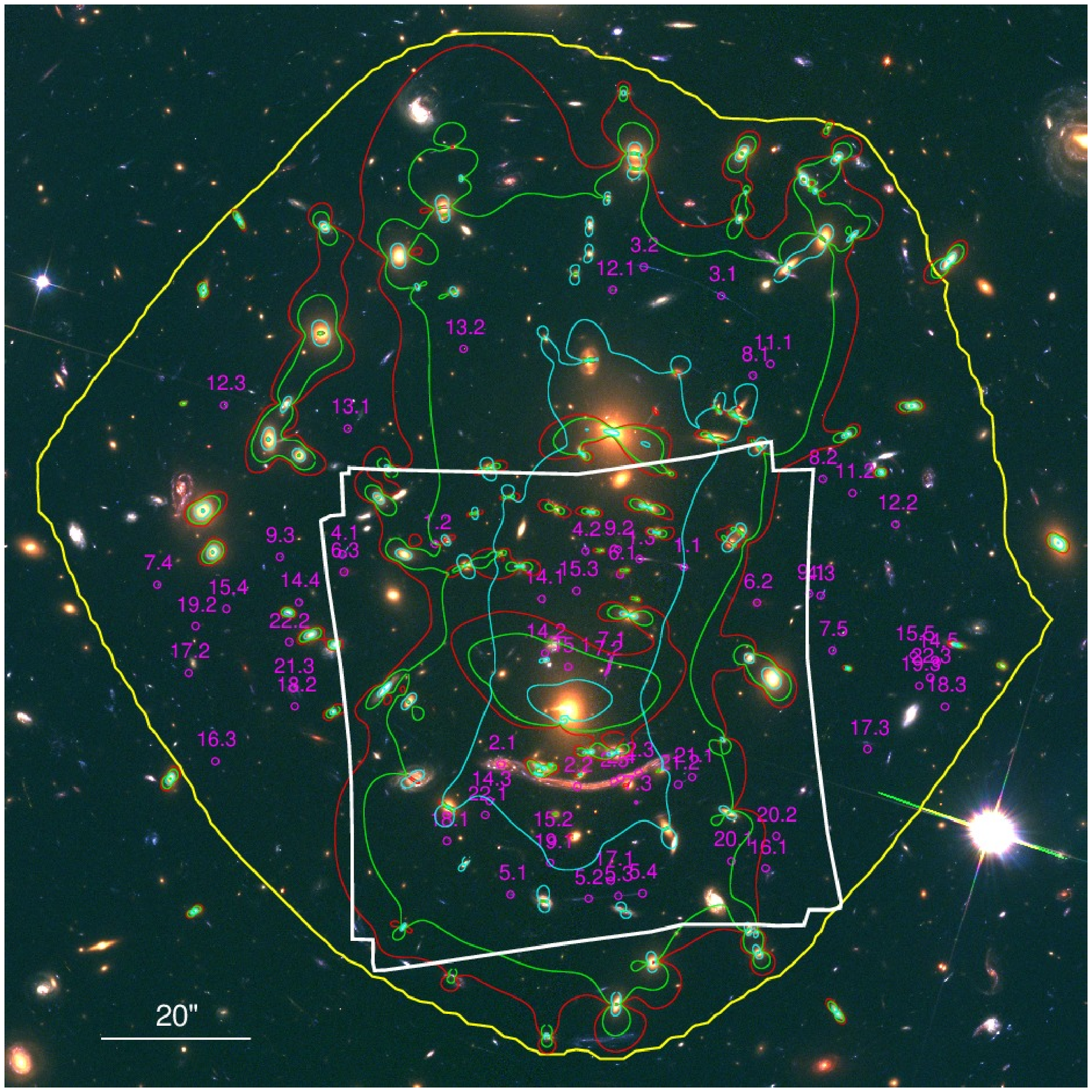}
\caption{Extent of the ``multi-image zone'' (yellow curve) as
  predicted by the best A370 mass model.  This region represents the
  area that encloses all expected multiple images out to high redshift
  ($z = 10$.)  The MUSE data footprint (white square) is again
  included for reference, and is considerably smaller than the
  multi-image zone.  Additionally, three critical curves are
  displayed, highlighting lines of high-magnification for three
  redshifts: $z = 1$ (cyan), $z = 3$ (green) and $z = 9$ (red).  These
  lines are heavily clustered around the BCGs and the northern galaxy
  crown.  Finally, all image constraints used in the model are shown
  as magenta circles. }
\label{fig:zCrit}  
\end{figure*}

Finally, we can compare our model (and all models) of A370 to other
clusters in the HFF program.  The combined rms error for our best fit
model ($\sigma_i$ = 0.94\arcsec) is higher than the error presented in
\citet{jon14} ($\sigma_i$ = 0.82\arcsec), comparable to the error in
\citet{ric14} ($\sigma_i$ = 0.93\arcsec), and smaller than the error
presented in \citet{ric10} ($\sigma_i$ = 1.76\arcsec).  However all of
these rms values are larger than the best fit models of most other HFF
clusters, such as MACS 0416 ($\sigma_i$ = 0.5\arcsec; \citealt{cam16})
or Abell 2744 ($\sigma_i$ = 0.7\arcsec; \citealt{jau16}).  Instead,
while not as large, our rms is more similar to MACS 0717 ($\sigma_i$ =
1.9\arcsec; \citealt{lim16}).  Like MACS 0717, A370 is a complex
cluster, with several interacting mass components covering a large
field of view.  In fact, the A370 ``multiple-image zone'' (Figure
\ref{fig:zCrit}), the region predicted to contain all multiple images
out to high redshift ($z = 10$), is the second largest of all of the
Frontier Fields, behind only MACS 0717.  Given the similarities, it is
therefore unsurprising that the typical rms is also high.

To reduce the total rms further, we need to explore the other areas of
the cluster with MUSE.  This will help to refine the model even more
and allow us to test whether or not our current interpretation is
correct.  In particular, the crown mass clump may only be a temporary
solution; future data may require additional large-scale halos, or
even suggest an alternative to the crown.  We cannot be sure of this
until we have additional constraints in the North.  Fortunately, such
a survey is currently underway (PI F. Bauer), and the results of this
study will be the topic of a future paper.

\section{Conclusions}
\label{End}
In this paper, we have presented a new mass model for the A370
cluster, taking advantage of deep HFF imaging and new MUSE
spectroscopy.  Our main conclusions are as follows:

\begin{itemize}
\item We present a MUSE-based redshift catalog for A370, consisting of
  120 secure redshifts ($\delta z < 0.1\%$), including 34
  multiply-imaged background objects (comprising 15 unique systems),
  13 singly-imaged background galaxies, 56 cluster members, 13
  foreground galaxies, and 4 stars.
  
\item Together, HFF and MUSE data are a powerful combination, greatly
  improving our ability to construct lens models.  MUSE spectroscopy
  is particularly valuable for this work, as it can be used to blindly
  identify new lensing constraints without selecting a specific
  redshift range.

\item After constructing a mass model with the new multiply-imaged
  constraints, we find two key differences with previous work:

  \subitem 1.) A central core that is flatter and less massive due to
  an increase in radial lensing systems

  \subitem 2.) The need for additional large-scale mass clumps (``the
  bar'' and ``the crown'') to better fit lensing constraints in the
  North of the cluster

\item A lack of model constraints in the North makes accurate
  comparisons difficult.  We will need more MUSE data, covering a
  larger area of the A370 cluster to discriminate between possible
  interpretations.  These data are currently being taken, and will be
  the subject of an upcoming paper.
\end{itemize}

Additionally, our model can be used as a guide for upcoming
observations, as high-magnification regions in our models (Figure
\ref{fig:zCrit}) should be ideal places to look for new, magnified
high-redshift galaxies.  With new MUSE spectroscopy on the way, the
A370 cluster should continue to provide a wealth of new information in
the future.

\section*{Acknowledgments}
 DJL, JR, BC, GM, VP and JM acknowledge support from the ERC starting
 grant CALENDS.  This work has been carried out thanks to the support
 of the ANR FOGHAR (ANR-13-BS05-0010-02) and the OCEVU Labex (ANR-
 11-LABX-0060). LW acknowledges support by the Competitive Fund of the
 Leibniz Association through grant SAW-2015-AIP-2.







\bsp	
\label{lastpage}
\end{document}